\documentclass[twocolumn,prb]{revtex4}

\usepackage{graphics}

\usepackage{epsfig} 

\usepackage{dcolumn}

\usepackage{amsmath}

\begin{document}

\title{Ion exchange phase transitions in ``doped'' water--filled  channels}

\author{ J.~Zhang$^1$, A.~Kamenev$^1$, B.~I.~Shklovskii$^{1,2}$}

\affiliation{ $^{1}$ Department of Physics, University of
Minnesota,
Minneapolis, MN 55455, USA\\
$^{2}$ William I. Fine Theoretical Physics Institute, University
of Minnesota, Minneapolis, MN 55455, USA}

\date{\today}

\begin{abstract}

Ion transport through narrow water--filled channels is impeded by
a high electrostatic barrier. The latter originates from the large
ratio of the dielectric constants of the water and a surrounding
media. We show that ``doping'', i.e. immobile charges attached to
the walls of the channel, substantially reduces the barrier. This
explains why most of the biological ion channels are ``doped''. We
show that at  rather generic conditions the channels may undergo
ion exchange phase transitions (typically of the first order).
Upon  such a transition a finite latent concentration of ions may
either enter or leave the channel, or be exchanged between the
ions of different valences. We discuss possible implications of
these transitions for the Ca-vs.-Na selectivity of biological Ca
channels. We also show that transport of divalent Ca ions is
assisted by their fractionalization into two separate excitations.

\end{abstract}

\maketitle

\maketitle

\section{Introduction}
\label{secintro}

\begin{figure}[ht]
\begin{center}
\includegraphics[height=0.24\textheight]{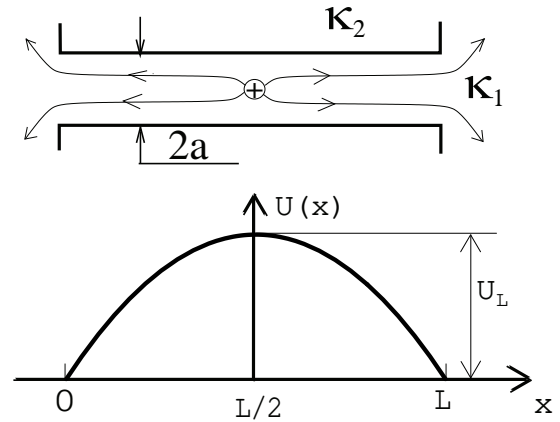}
\end{center}
\caption{ Electric field of a cation in a  cylindrical channel
with the large dielectric constant $\kappa_1\gg \kappa_2$. $L$ is
the channel length, $a$ is its radius. The self--energy barrier is
shown as a function of the $x$ coordinate. }
\label{figshortchannel}
\end{figure}

Protein ion channels functioning in biological lipid membranes is
a major frontier of biophysics ~\cite{Hille,Doyle}. An ion channel
can be inserted in an artificial membrane in vitro and studied
with  physical methods. For example, one can measure a
current--voltage response of a single water--filled channel
connecting two water reservoirs  (Fig.~\ref{figshortchannel}) as
function of concentration of various salts in the bulk. It is
well known~\cite{Parsegian,Jordan,Teber,Kamenev} that a neutral
channel creates a high electrostatic self--energy barrier for  ion
transport. The reason for this phenomena lies in the high ratio of
the dielectric constants of water,  $\kappa_1 \simeq 80$, and the
surrounding lipids, $\kappa_2\simeq 2$. For narrow channels the
barrier significantly exceeds $k_BT$ and thus constitutes a
serious impediment for ion transport across the membrane. It is a
fascinating problem to understand  mechanisms ``employed'' by
nature to overcome the barrier.

At a large concentration of salt in the surrounding water the
barrier can be suppressed by screening. However, for  biological
salt concentrations and narrow channels the screening is too weak
for that \cite{Kamenev}. As a result, at the ambient salt
concentrations even the screened barrier is usually well above
$k_BT$. What seems to be the ``mechanism of choice'' in narrow
protein channels is  ``{\em doping}''. Namely, there is a number
of amino-acids containing charged radicals. Amino-acids with, say,
negative charge are placed along the inner walls of the channels.
The static charged radicals are neutralized by the mobile cations
coming from the water solution. This provides a necessary high
concentration of mobile ions within the channel to suppress the
barrier~\cite{Zhang}.

Similar physics is at work in some artificial devices. For
example, water filled nanopores are studied  in silicon or silicon
oxide films~\cite{Li}. Dielectric constant of silicon oxide is
close to $4\ll 80$, so a very narrow and long artificial channels
may have large self-energy barrier. Ion transport through such a
channel can be facilitated by naturally appearing or intentionally
introduced wall charges, "dopants". Their concentration may be
tuned by pH of the solution.

The aim of this paper is to point out that the doping may lead to
another remarkable  phenomena: the ion exchange phase transitions.
An example of such a transition is provided by a negatively doped
channel in the solution of monovalent and divalent cations. At
small concentration of divalent cations, every dopant  is
neutralized by a single monovalent cation. If the concentration of
divalent salt increases above a certain critical concentration the
monovalent cations leave the channel, while divalent ones  enter
to preserve the charge neutrality. Since neutralization with the
divalent ions requires  only half as many ions, it may be carried
out with lesser entropy loss than the monovalent neutralization.
The specifics of the 1d geometry is that this competition leads to
the first order phase transition rather than a crossover (as is
the case for neutralizing 2d charged surfaces in the solution). We
show that the doped channels exhibit  rich phase diagrams in the
space of salt and dopant concentrations.

Let us remind the origin of the electrostatic self--energy
barrier. Consider a single cation placed in the middle of the
channel with the length $L$ and the radius $a$,
Fig.~\ref{figshortchannel}. If the ratio of the dielectric
constants is large $\kappa_1/\kappa_2\gg 1$, the electric
displacement $D$ is confined within the channel. As a result, the
electric field lines of a charge are forced  to propagate within
the channel until its mouth. According to the Gauss theorem the
electric field at a distance $x > a$ from the cation is uniform
and  is given by $E_0 = 2e/(\kappa_1 a^{2})$. The energy of such a
field in the volume of the channel is:
\begin{equation}
U_L(0) = {\kappa_1 E_{0}^{2}\pi a^{2}L \over 8\pi} = {e^{2}L \over
2\kappa_1 a^{\, 2}}={eE_{0}L \over 4}\, , \label{short}
\end{equation}
where the zero argument is added to indicate that there are no
other charges in the channel. The bare barrier, $U_{L}(0)$, is
proportional to $L$ and (for a narrow channel) can be much larger
than $k_BT$, making the channel resistance exponentially large.
\begin{figure}[ht]
\begin{center}
\includegraphics[height=0.065\textheight]{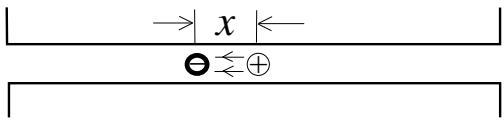}
\end{center}
\caption{A cation (in thin circle) bound to a negative wall charge
(in thick circle). When the cation moves away from the host the
energy grows linearly with the separation $|x|$. }\label{figpair}
\end{figure}
If a dopant with  the unit negative charge is attached to the
inner wall of the channel it attracts a mobile cation from the
salt solution (Fig.~\ref{figpair}). There is a confining
interaction potential $\Phi(x)= E_0|x|$ between them.  Condition
$e\Phi(x_T)=k_BT$ defines the characteristic thermal length of
such classical ``atom'', $x_T = k_BT/eE_0 = a^{2}/2l_{B}$, where
$l_B \equiv e^{2}/(\kappa_{1}k_{B}T)$ is the Bjerrum length (for
water at the room temperature $l_B=0.7\,$nm). This ``atom'' is
similar to an acceptor in a semiconductor (the classical length
$x_T$ plays the role of the effective acceptor Bohr radius). It is
convenient to measure the one--dimensional concentrations of both
mobile salt and dopants in units of $1/x_T$. In such units the
small concentration corresponds to non--overlapping neutral pairs
(``atoms''), while the large one corresponds to the dense plasma
of mobile and immobile charges. These phases are similar to
lightly and heavily doped $p$-type semiconductors, respectively.

It is important to notice that in both limits {\em all} the
charges interact with each other through the 1d Coulomb potential
\begin{equation}
\Phi(x_i-x_j)=\sigma_i\sigma_j  E_0|x_i-x_j|, \label{longrangepot}
\end{equation}
where $x_i$ and $\sigma_i=\pm 1$ are coordinates and charges of
both dissociated ions and dopants. Another way to formulate the
same statement is to notice that the electric field experiences
the jump of $2E_0\sigma_i$ at the location of the charge
$\sigma_i$. Because all the charges inside the channel are
integers in unit of $e$, the electric field is {\em conserved}
modulo $2E_0$. It is thus convenient to define the {\em order
parameter} $q\equiv \mbox{frac}[E(x)/2E_0]$, which is the same at
every point along the channel. The physical meaning of $q\in
[0,1]$ is the image charge induced in the bulk solution to
terminate the electric field lines leaving the channel. One may
notice that the adiabatic transport of a unit charge across the
channel is always associated with $q$ spanning  the interval
$[0,1]$. Indeed, a charge at a distance $x$  from one end of the
channel produces the fields $2E_0x/L$ and $2E_0(x/L-1)$ to the
right and left of $x$, correspondingly. Therefore $q=x/L$
continuously spans the interval $[0,1]$ as the charge moves from
$x=0$ to $x=L$.

To calculate the transport barrier (as well as the thermodynamics)
of the channel one needs to know the free energy $F_q$ of the
channel as a function of the order parameter. The equilibrium
(referred below as the {\em ground state}) free energy corresponds
to the minimum of this function $F_{\min}$. Transport of charge
and thus varying $q$ within $[0,1]$ interval is associated with
passing through the maximum of the $F_q$ function $F_{\max}$.
Throughout this paper we shall refer to such a maxima as the {\em
saddle point} state. The transport barrier is given by the
difference between the saddle point  and the ground state free
energies: $U_L=F_{\max} - F_{\min}$. The equilibrium
concentrations of ions inside the channel are given by the
derivatives of $F_{\min}$ with respect to the corresponding
chemical potentials related to concentrations in the bulk
solution. We show below that the calculation of the partition
function of the channel may be mapped on a fictitious quantum
mechanical problem with the periodic potential.  The function
$F_q$ plays the role of the lowest Bloch band, where $q$ is mapped
onto the quasi-momentum. As a result, the entire information of
the thermodynamical as well as transport properties of the channel
may be obtained from the analytical or numerical diagonalization
of the  proper ``quantum'' operator.

For an infinitely long channel with the long range interaction
potential Eq.~(\ref{longrangepot}) we arrive at true phase
transitions, in spite of the one--dimensional nature of the
problem. However, at finite ratio of dielectric constants electric
field lines exit from the channel at the distance $\xi \simeq
a(\kappa_1/\kappa_2)^{1/2}\approx 6.8\,a$. As a result, the
potential Eq.~(\ref{longrangepot}) is truncated and phase
transitions are smeared by fluctuations even in the infinite
channel. In practice all the channels have finite length which
leads to an additional smearing. We shall discuss sharpness  of
such smeared transitions below.
\begin{figure}[ht]
\begin{center}
\includegraphics[height=0.07\textheight]{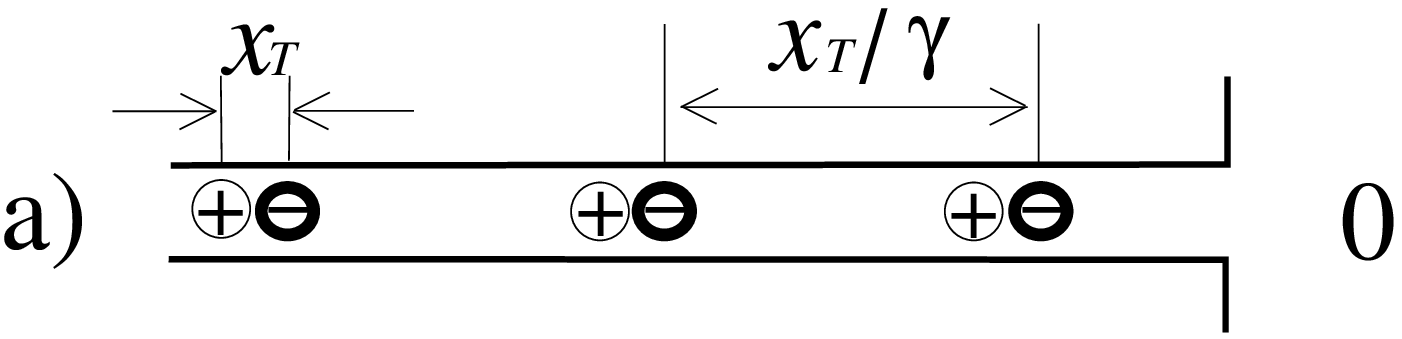} \hfill
\includegraphics[height=0.056\textheight]{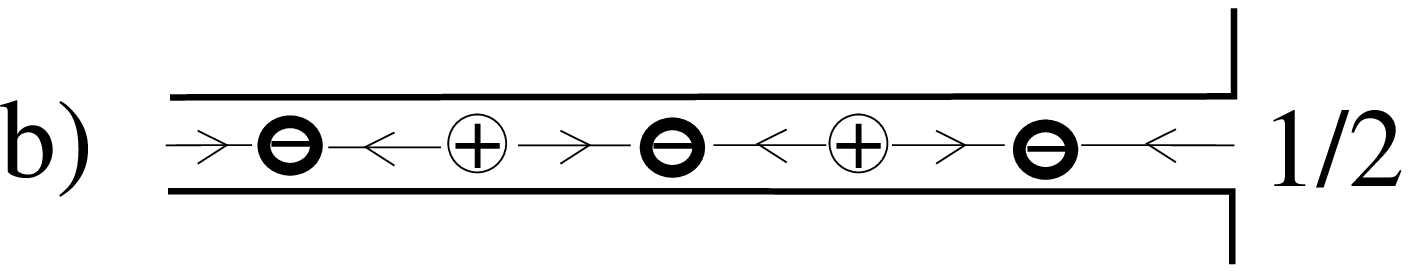}
\end{center}
\caption{The ground state and the transport saddle point of the
channel with negative monovalent dopants. Only the right part of
the channel is shown. (a) The ground state ($q=0$): all dopants
(thick circles)  bound mobile cations (thin circles). (b) The
transport saddle point ($q=1/2$): cations are free in  sections
between two adjacent dopants.}\label{figacceptor}
\end{figure}

The outline of this paper is as follows: In section \ref{sec2} we
briefly review results for the simplest model of periodically
placed negative charges in the monovalent salt solution (Fig.
\ref{figacceptor}), published earlier  in the short
communication~\cite{Zhang}. In this example the barrier is
gradually reduced with the increasing doping. Sections \ref{sec3}
-- \ref{sec5} are devoted to several modifications of the model
which, contrary to expectations raised by the results of section
\ref{sec2}, lead to ion exchange phase transitions. In section
\ref{sec3} we consider a channel with the alternating positive and
negative dopants in monovalent salt solution and study the phase
transition at which mobile ions leave the channel. In section
\ref{sec4} and \ref{sec5} we return to equidistant negative
dopants, but consider the role of divalent cations in the bulk
solution. In particular, in section \ref{sec4}, we assume that all
cations in the bulk solution are divalent and show that this does
{\em not} lead to four times increase of the self--energy barrier.
The reason is that the divalent ions are effectively
fractionalized in two monovalent excitations. In section
\ref{sec5} we deal with a mixture of monovalent and divalent
cations and study their exchange phase transition. We discuss
possible implications of this transition for understanding the
Ca-vs.-Na selectivity of biological Ca channels. For all the
examples we present transport barrier, latent ion concentration
and phase diagram along with the simple estimates, explaining the
observed phenomenology. The details of the mapping onto the
effective quantum mechanics as well as of the ensuing numerical
scheme are delegated to sections \ref{secanalytical} and
\ref{secnumerical}. Results of sections \ref{sec3} -- \ref{sec5}
are valid only for very long channels and true 1d Coulomb
potential.  In section \ref{secxi} we discuss the effects of
finite channel length and electric field leakage from the channel
on smearing of the phase transitions. In section \ref{Donnan} we
consider boundary effects at the channel ends leading to an
additional contact (Donnan) potential. We conclude in section
\ref{secconclusion} by brief discussion of possible
nano-engineering applications of the presented models.

\section{Negatively doped  channel in a monovalent solution }
\label{sec2}

As the simplest example of a doped channel we consider a  channel
with negative unit--charge dopants periodically attached to the
inner walls at distance $x_T/\gamma$ from each other (Fig.
\ref{figacceptor}). Here $\gamma$ is the dimensionless
one--dimensional concentration of dopants. At both ends (mouths)
the channel is in equilibrium with a monovalent salt solution with
the bulk concentration $c$. It is convenient to  introduce the
dimensionless monovalent salt concentration as $\alpha_1\equiv c
\pi a^2 x_T$. We shall restrict ourselves to the small salt
concentration (or narrow channels) such that $\alpha_1\ll 1$. In
this case the transport barrier of the undoped ($\gamma=0$)
channel is given by Eq.~(\ref{short}) (save for the small
screening reduction which scales as $1-4\alpha_1$,
[\onlinecite{Kamenev}]).

\begin{figure}[ht]
\begin{center}
\includegraphics[height=0.16\textheight]{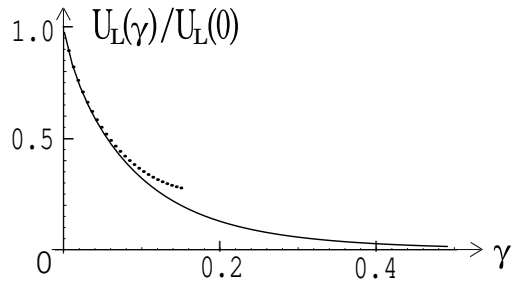}
\end{center}
\caption{The function $U_L(\gamma)/U_L(0)$ for $\alpha_1=10^{-5}$.
Its $\gamma \ll 1$ asymptotic, Eq.~(\ref{barriereg}), is shown by
the dotted line. } \label{figfgamma}
\end{figure}

The calculations, described in details in sections
\ref{secanalytical} and \ref{secnumerical}, lead to the barrier
plotted  in Fig.~\ref{figfgamma} as a function of the dopant
concentration $\gamma$. The barrier  decreases sharply as $\gamma$
increases. For example, a very modest concentration of dopants
$\gamma=0.2$ is enough to suppress the barrier more than five
times (typically bringing it below $k_B T$). There are {\em no}
phase transitions in this system in the entire phase space of
concentrations $\gamma$ and $\alpha_1$.

For the small dopant concentration, $\gamma\ll 1$, the result may
be understood with a simple reasoning. The ground state of the
channel corresponds to all negative dopants being locally
neutralized by the mobile cations from the solution
(Fig.~\ref{figacceptor} a). As a result, there is no  electric
field within the channel and thus in the ground state $q=0$. The
corresponding free energy has only entropic component (the absence
of the electric fields implies no energy cost) which is given by:
\begin{equation}\label{groundeg}
F_0 = -\gamma L k_B T \ln (2\alpha_1)= U_{L}(0) \cdot 4\gamma \ln
[1/(2\alpha_1)]\, .
\end{equation}
Indeed, bringing one cation from the bulk solution into the
channel to compensate a dopant charge leads to the entropy
reduction $S_0=k_B \ln (\pi a^2 2x_T c)=k_B\ln(2\alpha_1)$. Here
$\pi a^2  2x_T$ is the allowed volume of the cation's thermal
motion within the channel, while $c^{-1}$ is the accessible volume
in the bulk.

The maximum of the free energy is associated  with the state with
$q=1/2$, see Fig.~\ref{figacceptor} b. It can be viewed as a
result of putting a vacancy in the middle of the channel. The
latter creates the electric field $\pm E_{0}$ which orients the
dipole moments of all the ``atoms''. In other words, it orders all
the charges in an alternating sequence of positive and negative
ones. This unbinds the  cations from the dopants and makes them
free to move between neighboring dopants (Fig.~\ref{figacceptor}
b). Indeed, upon such rearrangement  the electric field is still
$E=\pm E_0$ everywhere in the channel, according to the Gauss
theorem.  Therefore, the energy of $q=1/2$ state is still given by
Eq.~(\ref{short}). However, its entropy is dramatically increased
with respect to $q=0$  state due to the unbinding of the cations:
the available volume is now $\pi a^2 x_T/\gamma$ and the resulting
entropy per cation is $S_{1/2}=k_B \ln (\alpha_1/\gamma)$. The
corresponding free energy of the saddle point state is:
\begin{equation}\label{saddleeg}
F_{1/2} = U_{L}(0) [1 - 4\gamma \ln (\alpha_1/\gamma)]\, .
\end{equation}
Recalling that the transport barrier is given by the difference
between the saddle point  and the ground state free energies, one
obtains:
\begin{equation}\label{barriereg}
U_L(\gamma)=U_{L}(0) [ 1 - 4\gamma \ln (1/2\gamma)]\, .
\end{equation}
This expression is plotted in  Fig.~\ref{figfgamma} by the dotted
line. It provides a perfect fit for the transport barrier at small
dopant concentration. Equation~(\ref{barriereg}) is applicable for
$\alpha_1 < \gamma \ll 1$. In the opposite limit $\gamma <\alpha_1
\ll 1$  more free ions may enter the channel in the saddle point
state. As a result, the calculation of $S_{1/2}$ should be
slightly modified \cite{Zhang}, leading to:
\begin{equation}\label{barrieracueg}
U_L(\gamma) = U_{L}(0) \left[1-4\gamma\ln\left({1\over
2\alpha_1}\sinh{\alpha_1\over \gamma}\right)\right] \, .
\end{equation}
This result, exhibiting  non--singular behavior in the small
concentration limit, is valid for an arbitrary relation between
$\alpha_1$ and $\gamma$ (both being small enough).

\section{Channel with alternating positive and negative dopants}
\label{sec3}

\begin{figure}[ht]
\begin{center}
\includegraphics[height=0.08\textheight]{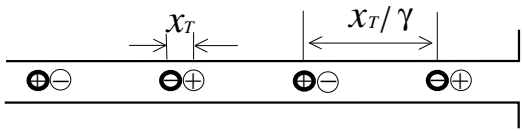}
\end{center}
\caption{A channel with the alternating dopants (thick circles).
The mobile counterions are shown as  thin circles. }
\label{figdopealternative}
\end{figure}

As a first example exhibiting the ion--exchange (actually
ion--release) phase--transition we consider a model of a
``compensated'' channel (the word compensation is used here by
analogy with  semiconductors, where acceptors can be compensated
by donors). This is the channel with positive and negative
unit--charge dopants alternating in one-dimensional NaCl type
lattice with the lattice constant $2 x_T/\gamma$
(Fig.~\ref{figdopealternative}).

\begin{figure}[ht]
\begin{center}
\includegraphics[height=0.16\textheight]{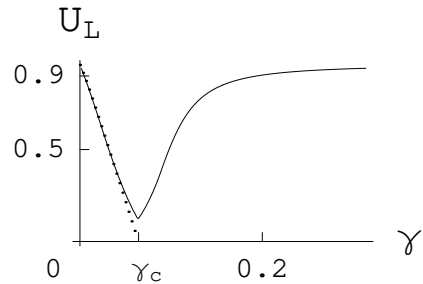}
\end{center}
\caption{ The transport barrier in units of $U_L(0)$ for the
``compensated'' channel of Fig.~\ref{figdopealternative} for
$\alpha_1=0.01$. The dotted line shows Eq.~(\ref{correctionb}).}
\label{figalternative}
\end{figure}

\begin{figure}[ht]
\begin{center}
\includegraphics[height=0.15\textheight]{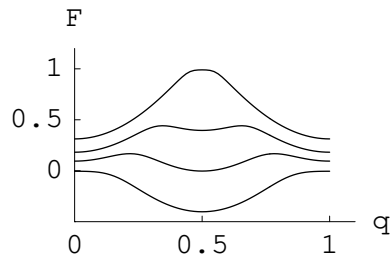}
\end{center}
\caption{Free energy in units of $U_L(0)$ as a function of $q$ for
$\gamma=0.02,~0.05,~0.07$ and $0.09$ (from top to bottom) at
$\alpha_1=0.01$. The lower three graphs are vertically offset for
clarity.} \label{qalternative}
\end{figure}

The channel is filled with the solution of the monovalent salt
with the bulk dimensionless concentration $\alpha_1=\pi a^2 x_T
c$. The transport barrier calculated for $\alpha_1=0.01$ as a
function of the dopant concentration $\gamma$ is depicted  in
Fig.~\ref{figalternative}.
One observes a characteristic sharp dip in a vicinity of a certain
dopant concentration $\gamma_c\approx 0.06$. To clarify a reason
for such an unexpected behavior (cf. Fig.~\ref{figfgamma}) we plot
the free energy as a function of the order parameter $q$ for a few
values of $\gamma$ close to $\gamma_c$, see
Fig.~\ref{qalternative}. Notice that for small $\gamma$ the
minimum of the free energy is at $q=0$, corresponding to the
absence of the  electric field inside the channel. The maximum is
at $q=1/2$, i.e. the electric field  $\pm E_0$. However, once the
dopant concentration $\gamma$ increases the second minimum
develops at $q=1/2$, which eventually overcomes the $q=0$ minimum
at $\gamma=\gamma_c$. In the limit of large $\gamma$ the ground
state corresponds to $q=1/2$ (electric field $\pm E_0$), while the
saddle point state is at $q=0$ (no electric field). It is clear
from Fig.~\ref{qalternative} that the transition between the two
limits is of the first order.

\begin{figure}[ht]
\begin{center}
\includegraphics[height=0.045\textheight]{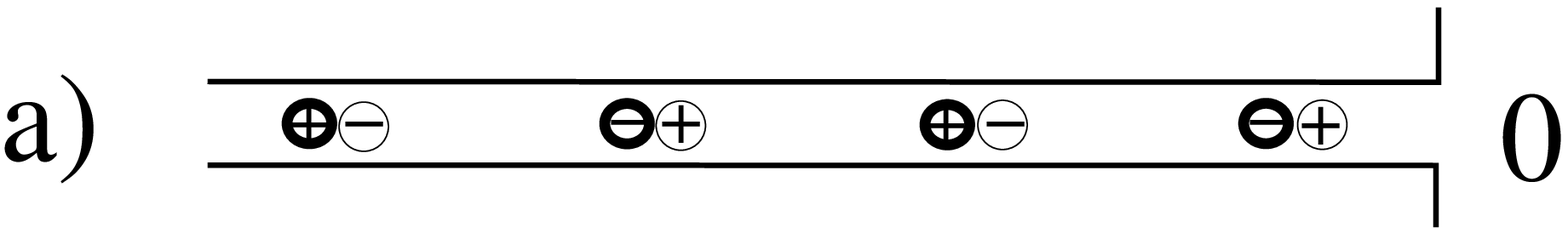} \hfill
\includegraphics[height=0.045\textheight]{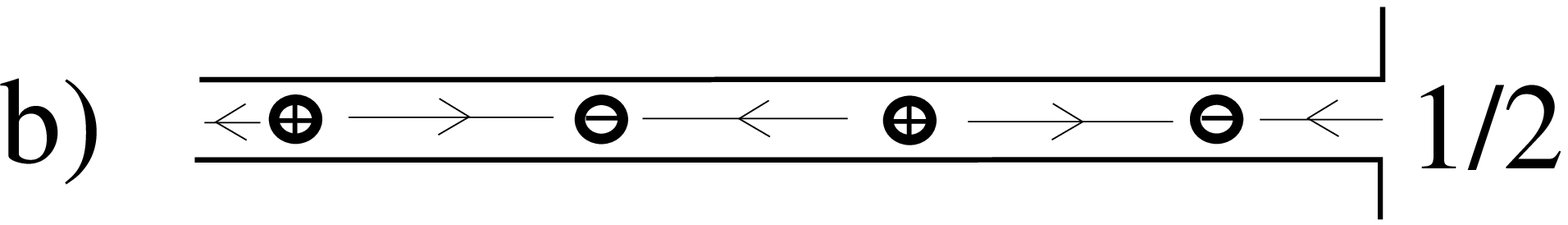} \hfill
\includegraphics[height=0.045\textheight]{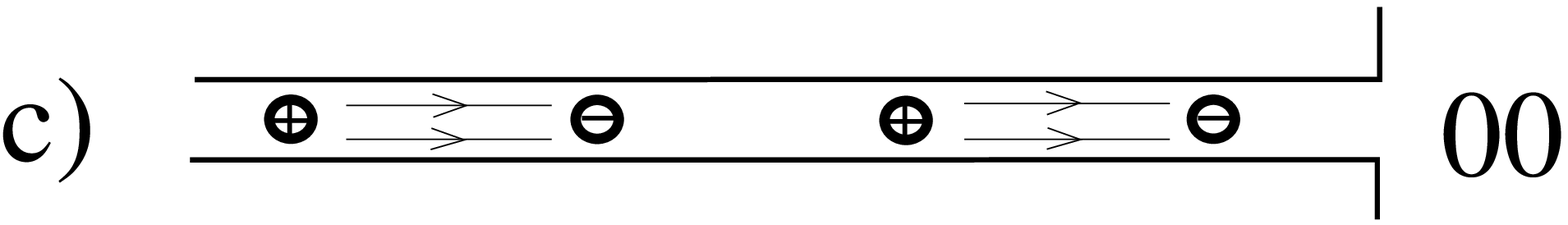}
\end{center}
\caption{States $0$, $1/2$ and $00$. The corresponding  free
energies are given by Eqs.~(\ref{ground1}), (\ref{ground2}) and
(\ref{line3}).}\label{figdopecompensate}
\end{figure}

To understand the nature of this transition, consider two
candidates for the ground state. The $q=0$ state, referred below
as $0$, is depicted on Fig.~\ref{figdopecompensate} a. In this
state every dopant tightly binds a counterion from the solution.
Such a state does not involve an energy cost and has the negative
entropy $S_0=k_B\ln(2\alpha_1)$ per dopant. As a result, the
corresponding free energy is (cf. Eq.~(\ref{groundeg})):
\begin{equation}\label{ground1}
F_0 = U_{L}(0) \cdot 4\gamma \ln [1/(2\alpha_1)]\, .
\end{equation}
An alternative ground state is that of the channel free from any
dissociated ions, Fig.~\ref{figdopecompensate} b. There is an
electric field $\pm E_0$ alternating between the dopants.  This is
$q=1/2$ state, or simply $1/2$ state. Since no mobile ions enter
the channel, there is no entropy lost in comparison with the bulk.
There is, however, energy cost for having electric field $\pm E_0$
across the entire channel. As a result, the free energy of the
$1/2$ state is (cf. Eq.~(\ref{short})):
\begin{equation}\label{ground2}
F_{1/2} = U_L(0)\, .
\end{equation}
Comparing Eqs.~(\ref{ground1}) and (\ref{ground2}), one expects
that the critical dopant concentration is given by
\begin{equation}\label{gammac}
\gamma_c = -\left[4 \ln(2 \alpha_1)\right]^{-1}\, .
\end{equation}
For $\gamma< \gamma_c$ the state $0$ is expected to be the ground
state, while for $\gamma>\gamma_c$ the state $1/2$  is preferable.

\begin{figure}[ht]
\begin{center}
\includegraphics[height=0.21\textheight]{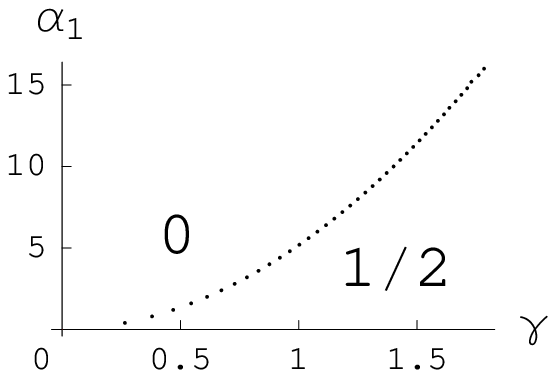} \hfill
\end{center}
\vspace{-5.2cm} \hspace{-1.5cm}
\includegraphics[height=0.1\textheight]{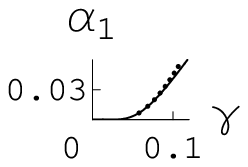}
\vspace{2.5cm} \caption{The phase diagram of the channel with
alternating doping (dotted lines). The phase boundary between the
$q=0$ and $q=1/2$ phases can be fitted as $\alpha_1\approx
5\gamma^2$ for $\alpha_1>1$. Inset: the small concentration part
of the phase boundary fitted by Eq.~(\ref{gammac}) (the full
line).} \label{figphase}
\end{figure}

In Fig.~\ref{figphase} we plot the phase diagram on the $(\gamma,
~\alpha_1)$ plane. The phase boundary between $0$ and $1/2$ states
is determined from the condition of having  two degenerate minima
of the free energy $F_q$ at $q=0$ and $q=1/2$. In the small
concentration limit (see the inset in Fig.~\ref{figphase}) the
phase boundary is indeed perfectly fitted by Eq.~(\ref{gammac}).
For larger concentrations it crosses over to $\gamma_c\propto
\sqrt{\alpha_1}$.  This can be understood as a result of a
competition between dopant separation $x_T/\gamma$ and the Debye
screening length $r_D\propto x_T/ \sqrt{\alpha_1}$. For $\gamma <
\sqrt{\alpha_1}$ we have $r_D < x_T/\gamma$, and thus each dopant
is screened locally by a cloud of mobile ions. As a result, the
neutral state $0$ is likely to be the  ground state. In the
opposite case $\gamma > \sqrt{\alpha_1}$ the  Debye length is
larger than the separation between oppositely charged dopants.
This is the limit of a week screening when most dopants may not
have counterions to screen them. Thus the $1/2$ state has a chance
to have a lower free energy.

\begin{figure}[ht]
\begin{center}
\includegraphics[height=0.17\textheight]{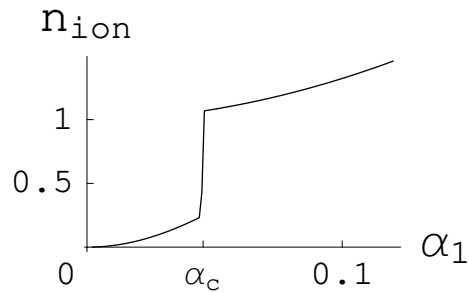}
\end{center}
\caption{The concentration of cations inside the channel in units
of $\gamma/x_T$ for $\gamma=0.1$. The discontinuous change occurs
at the  phase transition point.} \label{figlatentjump}
\end{figure}

\begin{figure}[ht]
\begin{center}
\includegraphics[height=0.17\textheight]{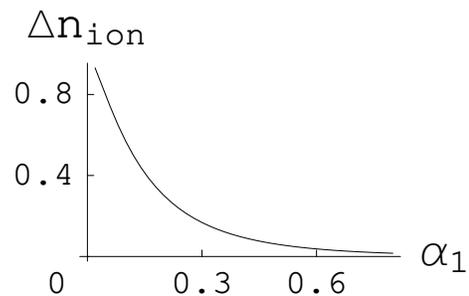}
\end{center}
\caption{The latent concentration of cations  in units of
$\gamma/x_T$ along the phase boundary line.} \label{figlatentnumb}
\end{figure}

Crossing the phase boundary in either direction is associated with
an abrupt change in the one-dimensional density of the salt ions
within the channel. The latter may be evaluated as the derivative
of the ground state free energy with respect to the chemical
potential of the salt: $n_{\mbox{ion}}= - {\alpha_1 \over L k_BT
}\, {\partial F_{\min}/ \partial \alpha_1}$. In
Fig.~\ref{figlatentjump} we plot concentration of  ions within the
channel, $n_{\mbox{ion}}$, in units of dopant concentration
$\gamma/x_T$ as a function of the bulk salt normality $\alpha_1$.
One clearly observes the latent concentration associated with the
first-order transition. As the bulk concentration increases past
the critical one, the mobile ions abruptly enter the channel. One
can monitor the latent concentration $\Delta n$ along the phase
transition line of Fig.~\ref{figphase}. In
Fig.~\ref{figlatentnumb} we plot the latent concentration along
the phase boundary as function of the critical $\alpha_1$. As
expected, in the dilute limit the latent concentration coincides
with the concentration of dopants (i.e. every dopant brings one
mobile ion). On the other hand, in the dense limit, the latent
concentration is exponentially small (but always finite).


Let us return to the calculation of the transport barrier,
Fig.~\ref{figalternative}. To this end one needs to understand the
nature of the saddle point states (in addition to that of the
ground states, discussed above). Deep in the phase where $0$ is
the ground state, the role of the saddle point state is played by
the $1/2$ state. Correspondingly  the transport barrier is
approximately given by the difference between Eq.~(\ref{ground2})
and Eq.~(\ref{ground1}). One may improve this estimate by taking
into account that in the $1/2$ state the free ions may enter the
channel (in even numbers to preserve the total charge neutrality).
This leads to the entropy of the $1/2$ state given by
$S_{1/2}=k_B\ln [ \sum_{k=0}^\infty (\alpha_1/\gamma)^{2k} /
(2k)!]$ per dopant. As a result, one finds for the transport
barrier in the $0$ state:
\begin{equation}
U_L(\alpha_1, \gamma) = U_L(0)
 \left[1-4\gamma\ln\left({1\over 2\alpha_1}\cosh{\alpha_1\over
\gamma}\right)\right]\, . \label{correctionb}
\end{equation}
This estimate is plotted in Fig.~\ref{figalternative} by the
dotted line.

\begin{figure}[ht]
\begin{center}
\includegraphics[height=0.19\textheight]{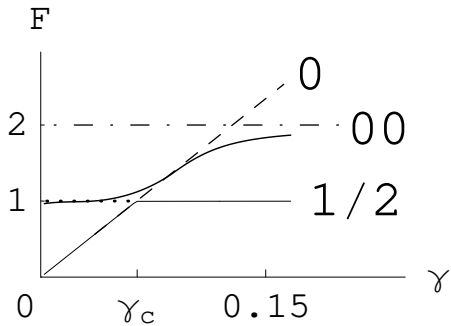}
\end{center}
\caption{Free energy diagram  for $\alpha_1=0.01$ as  function of
dopant concentration $\gamma$. The solid lines are numerical
results for the ground and saddle point states.
Eqs.~(\ref{ground1}), (\ref{ground2}) and (\ref{line3}),
describing states $0$, $1/2$ and $00$, correspondingly, are shown
by dashed, dotted and dash-dotted lines. } \label{figband}
\end{figure}

In the $1/2$ state, Fig.~\ref{figdopecompensate} b, the channel is
almost empty in the ground state configuration. The saddle point
may be achieved by putting a single cation in the middle of the
channel. This will rearrange the pattern of the internal electric
field as depicted in Fig.~\ref{figdopecompensate} c. This state
corresponds to $q=0$ and is denoted as $00$ (to distinguish it
from the state $0$, Fig.~\ref{figdopecompensate} a). Its free
energy (coinciding with the energy) is given by:
\begin{equation}\label{line3}
F_{00} = 2 U_L(0)\, .
\end{equation}
In Fig.~\ref{figband} we plot the free energies of the three
states $0$, $1/2$ and $00$ (cf. Fig.~\ref{figdopecompensate} and
Eqs.~(\ref{ground1}), (\ref{ground2}) and (\ref{line3})) as
functions of the dopant concentration $\gamma$. On the same graph
we also plot the calculated ground state free energy $F_{\min}$
along with the saddle point free energy $F_{\max}$. It is clear
that  the ground state undergoes  the first order transition
between $0$ and $1/2$ states at $\gamma=\gamma_c$. On the other
hand, the saddle point state experiences two smooth crossovers:
first between $1/2$ and $0$ and second between $0$ and $00$
states. The difference between the saddle point and the ground
state free energies is the transport barrier, which exhibits
exactly the type of behavior observed in
Fig.~\ref{figalternative}. Curiously, at large concentration of
dopants the barrier approaches exactly the same value as for the
undoped channel, $U_L(\alpha_1)$, [\onlinecite{Kamenev}]. This
could be expected, since extremely closely packed alternative
dopants compensate each other, effectively restoring the undoped
situation.

\section{Negatively doped channel with  divalent cations}
\label{sec4}

\begin{figure}[ht]
\begin{center}
\includegraphics[height=0.08\textheight]{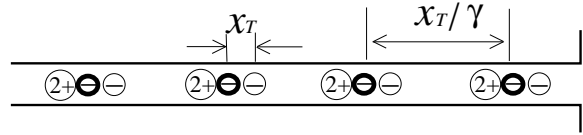}
\end{center}
\caption{A channel  with periodic dopants  (thick circles) and
divalent cations.  Mobile ions (Ca$^{2+}$ and Cl$^-$) are in thin
circles.} \label{figdopedivalent}
\end{figure}

In the above sections all the mobile ions as well as dopants were
monovalent. In this section we study the effect of cations being
{\em divalent} (e.g. Ca$^{2+}$, or Ba$^{2+}$), while  all negative
charges (both anions and dopants) are monovalent
(Fig.~\ref{figdopedivalent}). For example, one can imagine a
channel with negative wall charges in CaCl$_2$ solution. We denote
the dimensionless concentration of the divalent cations as
$\alpha_2$. The concentration of monovalent anions is simply
$\alpha_{-1}=2\alpha_2$.

\begin{figure}[ht]
\begin{center}
\includegraphics[height=0.22\textheight]{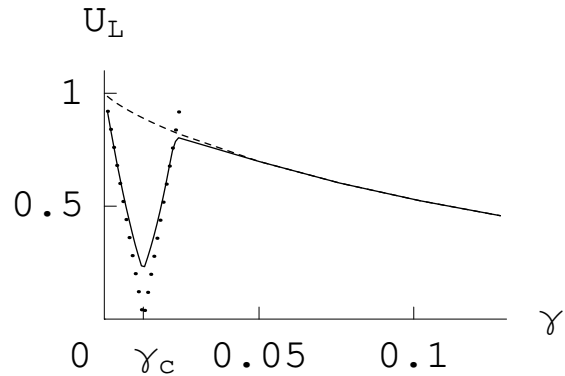}
\end{center}
\caption{The transport barrier  $U_L(\alpha_2,\gamma)$ in units of
$U_L(0)$ for $\alpha_2=5\cdot 10^{-7}$ (solid line). The dotted
line is $|F_{1/2}-F_0|$, and dashed line is $|F_{00}-F_{1/2}|$
calculated using Eqs.~(\ref{freeenergy0}), (\ref{freeenergy12})
and (\ref{freeenergy1}).} \label{figdoublefit}
\end{figure}

The transport  barrier as function of the dopant concentration
$\gamma$ for $\alpha_2=5\cdot 10^{-7}$ is shown in
Fig.~\ref{figdoublefit}.
Similarly to the case of the alternating doping (cf.
Fig.~\ref{figalternative}) the barrier experiences a sudden dip at
some critical dopant concentration $\gamma_c\approx 10^{-2}$. To
understand this behavior we looked at  the free energy $F_q$ as
function of the order parameter $q$ for several values of $\gamma$
in the vicinity of $\gamma_c$. The result is qualitatively similar
to that depicted in Fig.~\ref{qalternative}. Thus, it is again the
first order transition between two competing states that is
responsible for the behavior observed in Fig.~\ref{figdoublefit}.

\begin{figure}[ht]
\begin{center}
\includegraphics[height=0.046\textheight]{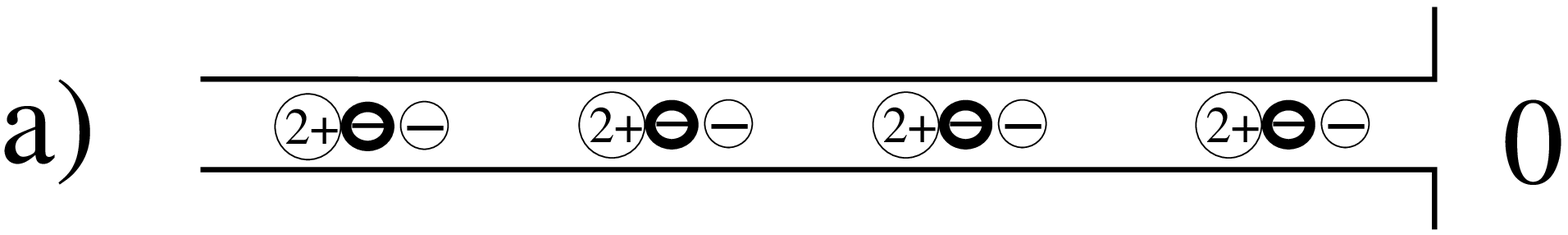} \hfill
\includegraphics[height=0.046\textheight]{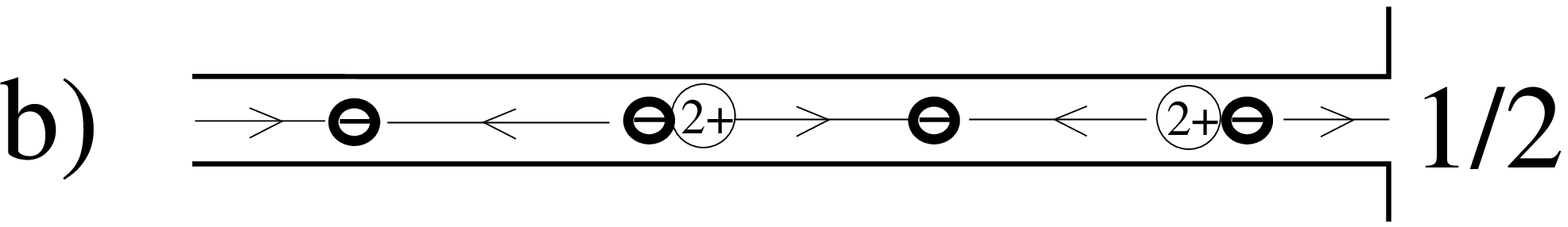} \hfill
\includegraphics[height=0.046\textheight]{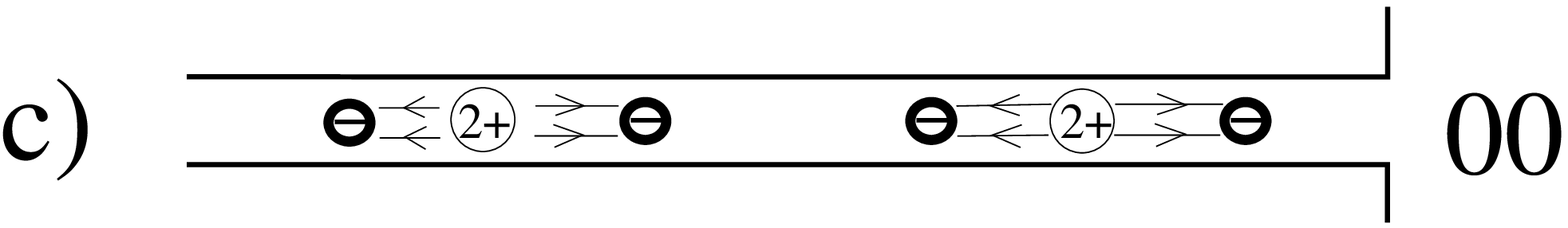}
\end{center}
\caption{The states $0$, $1/2$, and $00$  for the channel with
divalent cations. The corresponding free energies are given  by
Eqs.~(\ref{freeenergy0}), (\ref{freeenergy12}) and
(\ref{freeenergy1}). Dopants (thick circles) and anions (thin
circles) are monovalent, while cations (shown by $2+$ in thin
circles) are divalent. Electric fields are shown schematically,
one line per $E_0$. }\label{figdopedouble}
\end{figure}
The two candidates for the ground state, denoted, in accordance
with the fractional part of the internal electric field $q$, as
$0$ and $1/2$, are  depicted in Fig.~\ref{figdopedouble} a,b. In
the state $0$ every dopant is screened locally by an anion and a
doubly charged cation (Fig.~\ref{figdopedouble} a). The
corresponding free energy (consisting solely from the entropic
part) is given by~\cite{ft6}:
\begin{equation}\label{freeenergy0}
F_{0} = U_L(0)\cdot 4\gamma\ln\left({1 \over 6 \alpha_2^2}\right).
\end{equation}
The other state, $1/2$, has every second dopant overscreened by a
divalent cation (Fig.~\ref{figdopedouble} b). The free energy of
this state is:
\begin{equation}\label{freeenergy12}
F_{1/2} = U_L(0) \left[1+{4\gamma\over2}\ln(1 / \alpha_2)\right]\,
.
\end{equation}
Comparing Eqs.~(\ref{freeenergy0}) and (\ref{freeenergy12}), one
finds for the critical dopant concentration:
\begin{equation}\label{gammacdiv}
  \gamma_c=-\left[ 6\ln (\alpha_2) +4\ln6 \right]^{-1}\, .
\end{equation}
For $\alpha_2=5\cdot 10^{-7}$ it leads to $\gamma_c\approx 0.012$
in a good agreement with Fig.~\ref{figdoublefit}.

\begin{figure}[ht]
\begin{center}
\includegraphics[height=0.19\textheight]{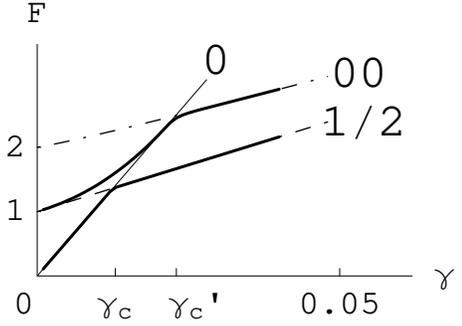}
\end{center}
\caption{Free energy diagram  for $\alpha_2=5\cdot 10^{-7}$ as a
function of dopant concentration $\gamma$. The full lines are
numerical results for the ground  and saddle point states
respectively. Eqs.~(\ref{freeenergy12}), (\ref{freeenergy1}) and
(\ref{freeenergy0}) describing states $1/2$, $00$ and $0$, are
shown by dashed, dash-dotted and solid thin line correspondingly.
State $0$ is the ground state for $\gamma<\gamma_c$, state $1/2$
is the ground state for $\gamma>\gamma_c$, and the state $00$ is
the saddle point for $\gamma>\gamma_c'$. The ground state
undergoes the first order phase transition at $\gamma_c$, while
the saddle point state evolves in a continuous way. }
\label{figdoubleband}
\end{figure}

To explain the transport barrier observed for small $\gamma$
(Fig.~\ref{figdoublefit}) one needs to know the saddle point
state. Such a state is depicted in Fig.~\ref{figdopedouble} c and
is denoted as $00$. It has one divalent cation trapped between
every other pair of dopants. It is easy to see that for such
arrangement the cations are free to move within the ``cage''
defined by the two neighboring dopants. This renders a rather
large entropy of the $00$ state. Its free energy is given by
\begin{equation}\label{freeenergy1}
F_{00} = U_L(0) \left[2+{4\gamma\over2}\ln({\gamma/
\alpha_2})\right]\, .
\end{equation}
In Fig.~\ref{figdoubleband} we plot the free energies of the
states $0$, $1/2$ and $00$, given by Eqs.~(\ref{freeenergy0}),
(\ref{freeenergy12}) and (\ref{freeenergy1}) correspondingly, as
functions of $\gamma$. On the same graph we plot calculated ground
state free energy $F_{\min}$ along with the saddle point free
energy $F_{\max}$ (full lines). One observes that the ground state
indeed undergoes the first order transition between the states $0$
and $1/2$ upon increasing $\gamma$ (lower full line). On the other
hand, the saddle point evolves smoothly from $1/2$ to $0$ and
eventually to $00$ (upper full line). The difference between the
two gives the transport barrier depicted in
Fig.~\ref{figdoublefit}, where  $|F_{1/2}-F_0|$ and
$|F_{00}-F_{1/2}|$ are also shown for comparison.

It is worth noticing  that on the both sides of the transition the
transport barrier  is close to $U_L(0)$, characteristic for the
transfer of the {\em unit} charge $e$. One could expect that for
charges $2e$ the self-energy barrier should be rather $4U_L(0)$.
This apparent reduction of the charge seems natural for the very
small $\gamma$. Indeed, the corresponding  ground state
(Fig.~\ref{figdopedouble} a) contains complex ions (CaCl)$^{1+}$
with the charge $e$. On the other hand, for $\gamma > \gamma_c$
the channel is free from Cl$^-$ ions and the current is provided
by Ca$^{2+}$  ions only. Thus, the observed barrier of $\leq
U_L(0)$ may be explained by fractionalization of charges $2e$ into
two charges $e$. To make such a fractionalization more transparent
one can redraw the saddle point state $00$ in a somewhat different
way, namely creating a soliton (domain wall, or defect) in the
state $1/2$.

\begin{figure}[ht]
\begin{center}
\includegraphics[height=0.066\textheight]{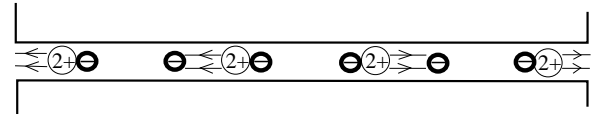}
\end{center}
\caption{The saddle point state $00$ represented as the unit
charge defect (soliton) within the ground state $1/2$. Ca$^{2+}$
ion added to the channel in the state $1/2$ fractionalizes in two
such solitons, each having an excess energy $\leq U_L(0)$. }
\label{figdoublemore}
\end{figure}

Fig.~\ref{figdoublemore} shows  the channel with a soliton in the
middle.
One can see that in this version of the $00$ state the fields
$2E_0$ and $0$  alternate similarly to Fig.~\ref{figdopedouble} c.
We can also see that  the soliton in the middle has the charge
$e$. Fractionalization of a single Ca$^{2+}$ ion in two
charge--$e$ defects  means that a Ca ion traverses the channel by
means of two solitons moving consecutively  across the channel.
The self-energy of each soliton (and therefore the transport
barrier) does not exceed $U_L(0)$.

\begin{figure}[ht]
\begin{center}
\includegraphics[height=0.2\textheight]{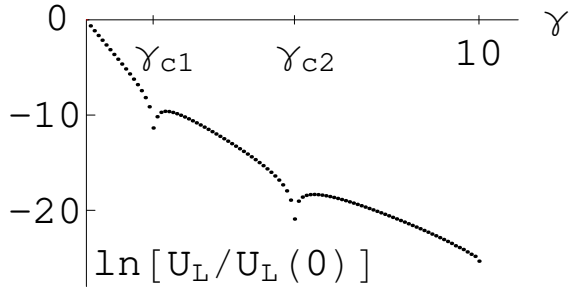}
\end{center}
\caption{ $\ln[U_L(\gamma)/U_L(0)]$ for $\alpha_2=5*10^{-5}$. Each
dip signals the presence of a phase transition.}
\label{figdoubleabs}
\end{figure}
\begin{figure}[ht]
\begin{center}
\includegraphics[height=0.2\textheight]{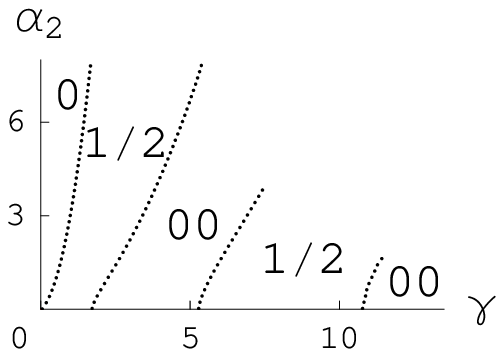} \hfill
\end{center}
\vspace{-5.0cm} \hspace{2.8cm}
\includegraphics[height=0.1\textheight]{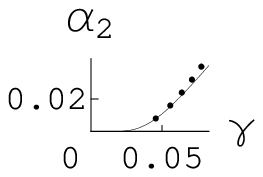}
\vspace{2.5cm} \caption{The phase diagram of the channel with the
divalent cations on the $(\gamma,~\alpha_2)$ plane. The dotted
lines are the boundaries between phases. The ground states in each
region are labelled. Inset: magnification  of the first phase
boundary line at small $\gamma$. The solid line is
Eq.~(\ref{gammacdiv}).} \label{figphases}
\end{figure}

So far the physics of the first order transition in the divalent
system was rather similar to that in the channel with the
alternating doping and a monovalent salt. There are, however,
important distinctions,  taking place at larger dopant
concentration. The logarithm of the transport barrier in the wide
range of $\gamma$ is plotted in Fig.~\ref{figdoubleabs}. One
notices a series of additional dips at $\gamma_{c1}\approx 1.7$,
$\gamma_{c2}\approx 5.3$, $\gamma_{c3}\approx 10.8$, etc. (the
first order transition at $\gamma_{c}\approx 0.01$, discussed
above, is not visible at this scale). These dips are indications
of the sequence of reentrant phase transitions, taking place at
larger $\gamma$. The calculated phase diagram  is plotted in
Fig.~\ref{figphases}. The leftmost phase  boundary line
corresponds to the first order phase transition between $0$ and
$1/2$ states. Its low concentration part along with the fit with
Eq.~(\ref{gammacdiv}) is magnified in the inset. The other lines
are transitions spotted in Fig.~(\ref{figdoubleabs}). We discuss
them in Appendix \ref{app1}.

\section{negatively doped Channel in solution with monovalent and
divalent cations}
\label{sec5}

We turn now to  the study of the channel in a solution with the
mixture of monovalent cations with the dimensionless concentration
$\alpha_1$ and divalent cations with the concentration $\alpha_2$.
Neutrality of the solution is maintained by monovalent anions with
the concentration $\alpha_{-1}=\alpha_1+2\alpha_2$. The channel is
assumed to be doped with the unit charge  negative dopants,
attached periodically  with the concentration $\gamma$.

\begin{figure}[ht]
\begin{center}
\includegraphics[height=0.19\textheight]{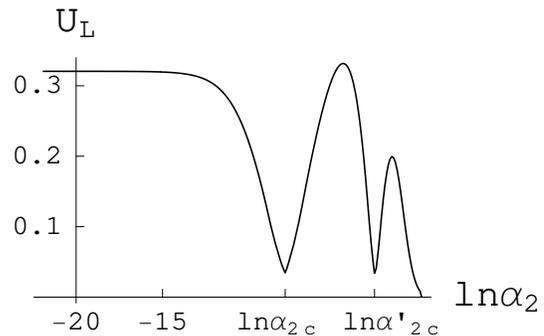}
\end{center}
\caption{The transport barrier as a function of $\ln \alpha_2$ for
$\gamma=0.1$ and $\alpha_1=0.001$.} \label{figselecbarrier}
\end{figure}

In Fig.~\ref{figselecbarrier} we plot the barrier as a function of
the  divalent cation concentration $\alpha_2$ for $\gamma=0.1$ and
$\alpha_1=0.001$.
The overall decrease of the barrier with the growing ion
concentration is interrupted by the two sharp dips  at
$\alpha_{2c}$ and $\alpha_{2c}'$. By plotting the $F_q$ function
for several $\alpha_2$ in the vicinity of $\alpha_{2c}$ and
$\alpha_{2c}'$, one observes that they correspond to the two
consecutive  first order transitions. As $\alpha_2$ increases the
system goes from $q=0$ phase into $q=1/2$ phase and eventually
back into $q=0$ phase.

\begin{figure}[ht]
\begin{center}
\includegraphics[height=0.19\textheight]{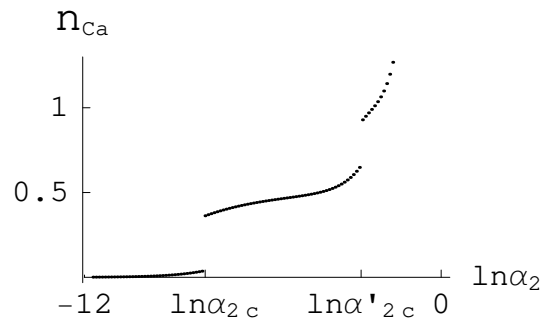}
\end{center}
\caption{Concentration of Ca$^{2+}$ ions in the channel in units
of $\gamma/x_T$ as function of $\ln \alpha_2$ for  $\gamma=0.1$
and $\alpha_1=0.001$. There are two discontinuous changes in Ca
concentration at  $\alpha_{2c}$ and $\alpha_{2c}'$. Each of these
two latent changes is close to the half of the number of dopants.
} \label{figsellatent}
\end{figure}

The concentration of divalent cations within the channel is given
by
\begin{equation}\label{number2}
n_{{Ca}}= -{\alpha_2 \over k_B T L} \, \left. { \partial F_{\min}
\over \partial \alpha_2} \right|_{\alpha_1, \alpha_{-1}}.
\end{equation}
It is plotted as a function of the bulk concentration in
Fig.~\ref{figsellatent}. One observes that at the first transition
the number of divalent cations entering the channel  is close to
the half of the number of dopants. When the second transition is
completed the number of divalent cations is approximately the same
as the number of dopants. This provides a clue on the nature of
the corresponding ground states. For $\alpha_2<\alpha_{2c}$ there
are almost no divalent cations in the channel. Therefore, both the
ground state and the saddle point state are the same as  in
sec.~\ref{sec2}, shown in Fig.~\ref{figacceptor}. The ground state
free energy and the transport barrier are given by
Eqs.~(\ref{groundeg}) and (\ref{barriereg}) correspondingly. At
$\alpha_2 =\alpha_{2c}$ the first order ion--exchange phase
transition takes place, where every two monovalent cations are
getting substituted by a single divalent one. The system's
behavior at larger $\alpha_2$ is qualitatively similar to that
described in the previous section. For
$\alpha_{2c}<\alpha_2<\alpha_{2c}'$ the ground state is the $ 1/2$
state, pictured in Fig.~\ref{figdopedouble} b. The corresponding
free energy is given by Eq.~(\ref{freeenergy12}). The second phase
transition at $\alpha_{2c}'$ is similar to that taking place on
the leftmost phase boundary  of Fig.~\ref{figphases} (which is
crossed now in the vertical direction). For
$\alpha_2>\alpha_{2c}'$ the ground state is the state $0$
(Fig.~\ref{figdopedouble} a) with the free energy given by
Eq.~(\ref{freeenergy0}).

\begin{figure}[ht]
\begin{center}
\includegraphics[height=0.19\textheight]{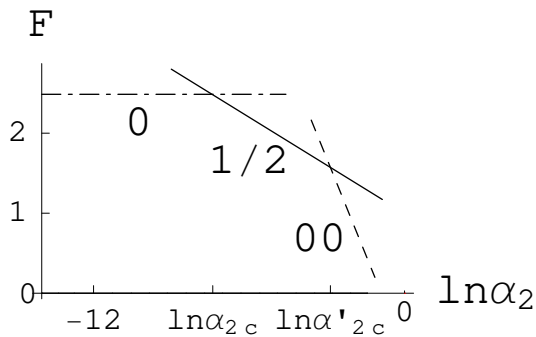}
\end{center}
\caption{Free energies of the three competing ground states for
$\gamma=0.1$ and $\alpha_1=0.001$. The dash-dotted line is
Eq.~(\ref{groundeg}), the solid line is Eq.~(\ref{freeenergy12}),
and the dashed line is Eq.~(\ref{freeenergy0}). The actual ground
state is chosen as the lowest of them.
 } \label{figselecenergy}
\end{figure}
The free energies of the three competing ground states as
functions of $\alpha_2$ are plotted in Fig.~\ref{figselecenergy}
for the same parameters as in Fig.~\ref{figselecbarrier}. They
indeed intersect at the concentrations close to $\alpha_{2c}$ and
$\alpha_{2c}'$. Since at each such intersection the symmetry of
the ground state (the $q$-value) changes, one expects that the
ground state changes via the first-order phase transition. The
critical value $\alpha_{2c}$ of the ion--exchange transition may
be estimated from Eqs.~(\ref{freeenergy12}) and (\ref{groundeg})
as:
\begin{equation}\label{crit}
\alpha_{2c} =(2\alpha_1)^2 e^{1/(2\gamma)} \, .
\end{equation}
Notice that it scales as $\alpha_1^2$ and therefore at small
concentrations the transition takes place at $\alpha_{2c}\ll
\alpha_1$. This is a manifestation of the law of mass action. The
second critical value may be estimated from Eq.~(\ref{gammacdiv})
as $\alpha_{2c}'=e^{-1/(6\gamma)}$ and is approximately
independent on $\alpha_1$. Comparing the two, one finds that the
transitions may take place only for small enough concentration of
the monovalent ions $\ln\alpha_1\lesssim -\gamma/3-\ln 2$. For
larger $\alpha_1$ there is a smooth crossover between the two
$q=0$ states.

\begin{figure}[ht]
\begin{center}
\includegraphics[height=0.27\textheight]{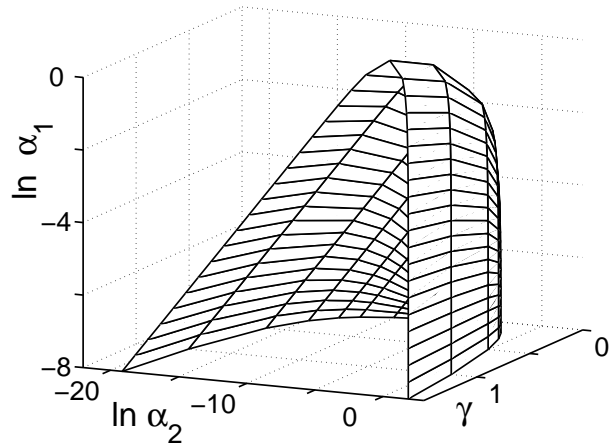}
\end{center}
\caption{Phase diagram in the space of cation and dopant
concentrations ($\ln \alpha_2-\gamma-\ln\alpha_1$). Inside the
tent-like shape  the ground state is $q=1/2$, outside -- the
ground state is $q=0$. } \label{figselecphase}
\end{figure}

The phase diagram in the space of cation and dopant concentrations
($\ln \alpha_2-\gamma-\ln\alpha_1$) is plotted  in
Fig.~\ref{figselecphase}. By fixing some $\gamma$ and not too
large $\alpha_1$ and varying $\alpha_2$ one crosses the phase
boundary twice. This way one observes two first order phase
transitions: from $0$ to $1/2$ and then from $1/2$ to $00$. The
corresponding transport barrier is shown in
Fig.~\ref{figselecbarrier}.

\begin{figure}[ht]
\begin{center}
\includegraphics[height=0.17\textheight]{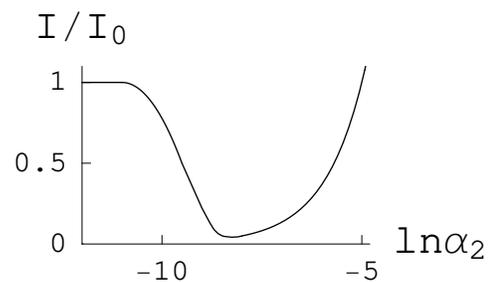}
\end{center}
\caption{Schematic plot  of the current through a Ca channel as a
function of Ca$^{2+}$ concentration, cf. Ref.~\onlinecite{Almers}.
The Na$^+$ concentration is $\alpha_1 \approx 10^{-2}$, the wall
charge concentration is $\gamma \approx 0.3$, and the (smeared)
transition is at $\ln \alpha_{2c}\approx -10$.}
\label{figseleccurrent}
\end{figure}

The model presented in this section is a simple cartoon for
Ca$^{2+}$ selective channels. Let us consider the total current
through the channel, $I$, equal to the sum of sodium and calcium
currents. Each of these partial currents in turn is determined by
two series resistances: the channel resistance and the combined
contact (mouth) resistances (the latter are inversely proportional
to the concentration of a given cation in the bulk). We assume
that at the biological concentration of Na$^{+}$ the current $I =
I_0$ and discuss predictions of our model of negatively doped
channel at $\gamma\approx 0.3$ regarding $I/I_0$ with growing
concentration $\alpha_2$ of Ca$^{2+}$ cations
(Fig.~\ref{figseleccurrent}). For $\alpha_2 < \alpha_{2c}$ the
channel is populated by Na$^+$ and Na$^+$ current dominates. For
$\alpha_2 > \alpha_{2c}$ the Na$^+$ current is blocked because
Na$^+$ ions are expelled  from the channel and substituted by the
Ca$^{2+}$. Indeed, a monovalent Na$^+$ cation cannot unbind the
divalent Ca$^{2+}$  from the dopants. Therefore, the transport
barrier for Na$^{+}$ ions is basically the bare barrier $U_L(0)
\gg k_BT$. In these conditions the Ca$^{2+}$ resistance of the
channel is small, because Ca$^{2+}$ concentration in the channel
is large. However, at the transition $\alpha_2 = \alpha_{2c}\ll
\alpha_1$ the concentration of Ca$^{2+}$ ions in the bulk water is
so small that the contact Ca$^{2+}$ resistance is very large.
Therefore, the Ca$^{2+}$ current is practically blocked and the
total current, $I$, drops sharply at $\alpha_{2c}$. As $\alpha_2$
grows the Ca$^{2+}$ current increases proportional to $\alpha_2$
due to decreasing contact resistance. As a result, one arrives at
the behavior of $I/I_0$ schematically shown in
Fig.~\ref{figseleccurrent}. This behavior is in a qualitative
agreement with the experimental data of
Refs.~[\onlinecite{Almers,Hille}].


\section{Analytical approach}
\label{secanalytical}

Consider a gas consisting of $N$ mobile monovalent cations  and
$N'$ mobile monovalent anions along with non--integer boundary
charges $q$ and $q'$ placed at $x=0$ and $x=L$ correspondingly. We
also consider a single negative unit dopant charge attached  at
the point $x_0$ inside the channel: $0<x_0<L$. The resulting
charge density takes the form:
\begin{equation}\label{chargedensity}
  \rho(x)\equiv \!\!\sum\limits_{j=1}^{N+N'}\!\! \sigma_j
  \delta(x-x_j)+q\delta(x)+q'\delta(x-L) - \delta(x-x_0)\, ,
\end{equation}
where $x_j$ stay for  coordinates of the mobile charges and
$\sigma_j=\pm 1$ for their charges. The interaction energy of such
a plasma is given by:
\begin{equation}\label{totalenergy_q}
U = {e\over 2}\int\!\!\!\int\limits_{0}^{L}\!\! dxdx'
\rho(x)\Phi(x-x')\rho(x')\, ,
\end{equation}
where the 1d Coulomb potential $\Phi(x)= \Phi(0) - E_0|x|$ is the
solution of the Poisson equation: $\nabla^2\Phi = -2E_0\delta(x)$.
(The self-energy $e\Phi(0)$ will be eventually taken to infinity
to enforce charge neutrality).

We are interested in the grand-canonical partition function of the
gas defined as
\begin{eqnarray}\label{partition5}
  && Z_L(q,q') = \sum\limits_{N,N'=0}^\infty e^{\mu(N+N')/(k_BT)}{1\over
  N!N'!} \\
&&\times \prod\limits_{j=1}^{N+N'}\left({\pi a^{\,2}\over
l_0^2}\int\limits_{0}^{L} {dx_j\over l_0} \right) e^{- U/(k_B T)}
\, ,\nonumber
\end{eqnarray}
where $\mu$ is the chemical potential (the same for cations and
anions) and $l_0$ is a microscopic scale related to the bulk salt
concentration as $c=e^{\mu/k_BT}/l_0^{3}$. Factor $\pi a^2/l_0^2$
originates from the integrations over transverse coordinates.

To proceed with the evaluation of $Z_L(q,q')$ we introduce the
resolution of unity written in the following way:
\begin{widetext}
\begin{eqnarray}\label{res_unity}
  1&=&\int\!\!{\cal D}\! \rho(x)\,\, \delta\!\!\left(\rho(x) -
\sum\limits_{j=1}^{N+N'} \sigma_j
  \delta(x-x_j)-q\delta(x)-q'\delta(x-L)+\delta(x-x_0) \right)
  \nonumber \\
  &=&\int\!\!\!\int{\cal D}\! \rho(x){\cal D} \theta(x)\,\,
   \exp\left\{\,-i\left(\int\limits_{0}^{L}  \!dx\, \theta(x)\rho(x)  -
\sum\limits_{j=1}^{N+N'} \sigma_j
  \theta(x_j)-q\theta(0)-q'\theta(L) + \theta(x_0) \right)\right\}~.\nonumber
\end{eqnarray}
Substituting this identity into Eq.~(\ref{partition5}), one
notices that the integrals over $x_j$ decouple and can be
performed independently. The result of such integration along with
the summation over $N$ and $N'$ is $\exp\{2\pi a^2 c\int_0^L dx
\cos\theta(x)\}$. Evaluation of the Gaussian integral over
$\rho(x)$ yields the exponent of
$\theta(x)\Phi^{-1}(x-x')\theta(x')$.  According to the Poisson
equation the inverse potential is $\Phi^{-1}(x-x')=-(2E_0)^{-1}
\delta(x-x') \partial^2_x$. As a result, one obtains for the
partition function:
\begin{eqnarray}\label{partition1}
  Z(q,q')= && \int\!\!\!\! \int\limits_{-\infty}^{\infty}
  \frac{d\theta_0
  d\theta_L d\theta_{x_0} }{(2\pi)^3}\,\, e^{ iq\theta_i+iq'\theta_f
-i\theta_{x_0} }\int\!\! {\cal D} \theta(x)\,\,
   \exp\left\{  -   \int\limits_0^L\!\! dx\left[{x_T\over 4} (\partial_x
\theta)^2 - {2\alpha_1\over
 x_T} \cos \theta(x)\right]    \right\} \, ,\nonumber
\end{eqnarray}
\end{widetext}
where $\alpha_1=\pi a^2 x_T c$.
 The integral over $\theta(x)$ runs over all
functions with the boundary conditions $\theta(0)=\theta_0\,$,
$\theta(L)=\theta_L$ and $\theta(x_0)=\theta_{x_0}$.

It is easy to see that this expression represents the matrix
element of the following T-exponent (or rather X-exponent)
operator:
\begin{equation}\label{Texponent}
  Z(q,q')=\langle q|e^{-{x_0\over x_T}\,  \hat H}
  e^{-i\theta} e^{-{L-x_0\over x_T} \, \hat H}|q'\rangle \, ,
\end{equation}
where the Hamiltonian is given by $\hat H=(i\hat\partial_\theta)^2
-2\alpha_1 \cos\theta$ and $|q\rangle$ is the eigenstate of the
momentum operator $i\hat\partial_\theta$. Since the Hamiltonian
conserves the momentum up to an integer value, $q'=q+M$, one can
restrict the  Hilbert space down to the subspace  with the fixed
{\em fractional} part of the boundary charge $0\leq q<1$. In this
subspace one can perform the gauge transformation, resulting in
the Mathieu Hamiltonian with the ``vector potential''
\cite{Lenard,Edwards,Kamenev}:
\begin{equation}\label{Mathieu}
\hat H_q =  (i\hat \partial_\theta -q)^2 -2\,\alpha_1 \cos
\theta\, .
\end{equation}
It acts in the space of periodic functions:
$\Psi(\theta)=\Psi(\theta+2\pi)$. Finally, taking the
``democratic'' sum over all integer parts of the boundary charge
(with the fixed fractional part $q$), one obtains:
\begin{equation}\label{partition-final}
Z(q)=\mbox{Tr}\left\{ e^{-\hat H_q{x_0\over x_T}}\,e^{-i\theta}\,
e^{-\hat H_q{L-x_0\over x_T}} \right\} \, .
\end{equation}

In the  more general situation the solution contains a set of ions
with charges (valences) $m\in {\cal Z}$ and the corresponding
dimensionless concentrations $\alpha_m$. The condition of total
electro-neutrality demands that:
\begin{equation}\label{neutrality}
  \sum\limits_m m\, \alpha_m =0 \, .
\end{equation}
The Mathieu Hamiltonian (\ref{Mathieu}) should be generalized as
\cite{Edwards}:
\begin{equation}
                                   \label{Mathieu-gen}
\hat H_q =  (i\hat \partial_\theta -q)^2 -\sum\limits_m\alpha_m\,
e^{im\theta} \, .
\end{equation}
Despite of being non--Hermitian, this Hamiltonian still possesses
a real band--structure~\cite{ftnt} $\epsilon_q^{(j)}$.
It is safe to assume that monovalent anions are always present in
the solution: $\alpha_{-1}>0$. This guarantees that the
band--structure has the unit period in $q$.

Consider first an undoped channel. Its partition function is given
by $Z(q)=\mbox{Tr}\{\exp(-\hat H_qL/x_T)\}$. In the long channel
limit, $L\gg x_T$, only the ground state, $\epsilon^{(0)}_q$, of
the Hamiltonian (\ref{Mathieu}) or (\ref{Mathieu-gen}) contributes
to the partition function. As a result, the free energy of the 1d
plasma is given by
\begin{equation}
      \label{free-energy}
F_q=k_BT \epsilon^{(0)}_q(\alpha_m) L/x_T=4U_L(0)
\epsilon^{(0)}_q(\alpha_m)  \, .
\end{equation}
The equilibrium ground state of the plasma corresponds to the
minimal value of the free energy. For the Mathieu Hamiltonian
(\ref{Mathieu}) $\epsilon^{(0)}_q$ has a single minimum  at $q=0$
(no induced charge at the boundary). It is also the case for the
more general Hamiltonians (\ref{Mathieu-gen}). Therefore the
equilibrium state of a neutral 1d Coulomb plasma does {\em not}
have a dipole moment and possesses  the reflection symmetry.

The adiabatic transfer of the unit charge across the channel is
associated with the slow change of $q$ from $q=0$ to $q=\pm  1$.
In this way the system must overcome the free energy maximum at
some value of $q$ (for the operators (\ref{Mathieu}) and
(\ref{Mathieu-gen})  the maximum  is at $q = 1/2$). As a result,
the activation barrier for the charge transfer is proportional to
the band-width of the lowest Bloch band \cite{Kamenev}:
\begin{equation}
                                         \label{barrier}
U_L(\alpha_m) = 4 U_L(0)
\left(\epsilon^{(0)}_{\mbox{max}}(\alpha_m) -
\epsilon^{(0)}_{\mbox{min}}(\alpha_m)\right) \, .
\end{equation}
Notice that in the ideal 1d Coulomb plasma the transport barrier
scales as the system size. It is also worth mentioning that both
the equilibrium free energy $F_0= 4 U_L(0)
\epsilon^{(0)}_{\mbox{min}}(\alpha_m)$ and the transport barrier,
Eq.~(\ref{barrier}), are smooth analytic functions of the
concentrations $\alpha_m$. Since there is the unique minimum and
maximum  within the interval $0\leq q<1$, there are {\em no} phase
transitions in the undoped channels.

Most of  biological ion channels have internal ``doping'' wall
charges within  the channel. If integer charges $n_1, n_2,\ldots,
n_N$ are fixed along the channel at the coordinates
$0<x_1<x_2<\ldots< x_N<L$,  the partition function is obtained by
straightforward generalization of Eq.~(\ref{partition-final}):
\begin{equation}\label{dopedpartition}
Z(q) =\mbox{Tr}\left\{ e^{-\hat H {x_1\over x_T} } e^{in_1\theta}
e^{-\hat H{x_2-x_1\over x_T} }\ldots e^{in_N\theta} e^{-\hat H{
L-x_N\over x_T} } \right\}\, .
\end{equation}
As long as all $n_k$ are integer, the boundary charge $q$ is a
good quantum number of the  operator under the trace sign. As a
result, the partition function is again a periodic function of $q$
with the unit period.

For the sake of illustration we shall focus on systems with
periodic arrangements of the wall charges. In this case the
partition function (\ref{dopedpartition}) takes the form:
$Z(q)=\mbox{Tr}\{\left( \hat {\cal U}_q \right)^N \}$, where $N$
is the number of dopants in the channel and $\hat {\cal U}_q$ is
the single--period evolution operator. We shall define the
spectrum of this operator as:
\begin{equation}
            \label{spectrum}
\hat {\cal U}_q\, \Psi^{(j)}_q(\theta) =
e^{-\epsilon_q^{(j)}/\gamma}\, \Psi^{(j)}_q(\theta) \, ,
\end{equation}
where $\gamma$ is the dimensionless concentration of dopants,
defined as $\gamma\equiv x_TN/L$. The evolution operator $\hat
{\cal U}_q$ is non--Hermitian. Its spectrum is nevertheless real
and symmetric function of $q$. The proof of this statement may be
constructed in the same way as for the operator in
Eq.~(\ref{Mathieu-gen}) \cite{ftnt}. The free energy of a long
doped system is given by $F_q =k_BT\epsilon_q^{(0)}N/\gamma = 4
U_L(0) \epsilon_q^{(0)} $ (cf. Eq.~(\ref{free-energy})). The
equilibrium ground state is given by the absolute minimum of this
function. Similarly, the transport barrier is given by
Eq.~(\ref{barrier}).

The simplest example is the periodic sequence of  unit--charge
negative dopants ($x_{k+1}-x_k=L/N$ and  $n_k=-1$) in the
monovalent salt solution,  section \ref{sec2}. The single--period
evolution operator takes the form:
\begin{equation}
                                 \label{one-period}
 \hat {\cal U}_q = e^{-i\theta} e^{-\hat H_q/\gamma } \, ,
\end{equation}
with $\hat H_q$ given by Eq.~(\ref{Mathieu}). As shown in
Ref.~\onlinecite{Zhang} its ground state
$\epsilon_q^{(0)}(\alpha_1,\gamma)$ is a function qualitatively
similar to the lowest Bloch band of the Mathieu operator in
Eq.~(\ref{Mathieu}). As a result, both equilibrium free energy and
the transport barrier are smooth function of the salt
concentration $\alpha_1$ and the dopant concentration $\gamma$.

The examples of sections \ref{sec4} and \ref{sec5} are described
by the evolution operators, which have the form of
Eq.~(\ref{one-period}) with the generalized Hamiltonian,
Eq.~(\ref{Mathieu-gen}). In the example of section \ref{sec4}
there are two non-zero concentrations: $\alpha_{-1}=2\alpha_2$,
while in section \ref{sec5} one deals with three types of ions:
$\alpha_{-1}=\alpha_1+2\alpha_2$. Finally, the alternating doping
example of section \ref{sec3} is described by the evolution
operator of the form
\begin{equation}
                                 \label{uq2}
 \hat {\cal U}_q = e^{-i\theta} e^{-\hat H_q/\gamma } e^{+i\theta}
e^{-\hat H_q/\gamma }
\end{equation}
with the Hamiltonian~(\ref{Mathieu}). Such operators may have more
complicated structure of their lowest Bloch band. In particular,
the latter may have more than one minimum within the period of the
reciprocal lattice: $q\in [0,1]$. The competition between (and
splitting of) the minima results in the first (second) order phase
transitions.

The analytic investigation of the spectrum of the  evolution
operators, such as Eq.~(\ref{one-period}), is possible in the
limits of small and large dopant concentrations $\gamma$. For
$\gamma\ll 1$ it follows from Eq.~(\ref{one-period}) that only the
lowest eigenvalues of $\hat H_q$ are important. It is then enough
to keep only the ground state of $\hat H_q$, save for an immediate
vicinity of $q=1/2$, where the ground and the first excited states
may be nearly degenerate.  If also $\alpha_m\ll 1$ the spectrum of
$\hat H_q$ along with the matrix elements of $e^{\pm i\theta}$ may
be calculated in the perturbation theory. Such  calculations  lead
to the free energies of the ground and saddle point states, which
are identical to those derived in sections \ref{sec2}--\ref{sec5}
using simple energy and entropy counting.

In the limit $\gamma\gg 1,\alpha_m$ one may develop a variant of
the  WKB approximation in the plane of the complex $\theta$,
\cite{Zhang}. It shows that the transport barrier and latent
concentration scales as $\exp\{-c\sqrt{\gamma}\}$, where
non-universal numbers $c$ are given by certain contour integrals
in the complex plane of $\theta$. In the example of section
\ref{sec2} we succeeded in quantitative prediction of the
coefficient $c$, see Ref.~[\onlinecite{Zhang}]. In the case of the
section \ref{sec4} the coefficient in front of $\cos(2\pi q)$ (see
Eq.~(\ref{doublecos})) is an oscillatory function of $\gamma$.
This probably translates into the complex value of the
corresponding $c$--constant. We did not succeed, however, in its
analytical evaluation.

\section{Numerical calculations}
\label{secnumerical}

By far the simplest way  to find  the spectrum of $\hat {\cal
U}_q$ is numerical. In the basis of the angular momentum $e^{ik
\theta}$ the Hamiltonian (\ref{Mathieu-gen}) takes the form of the
matrix:
\begin{equation}
                           \label{hamiltonianmatrix}
\left[\hat H_q\right]_{k,k'} = \left[ (k+q)^2
\delta_{k,k'}-\sum\limits_m \alpha_m \delta_{k,k'+m}\right]\, .
\end{equation}
In the same basis the dopant charge $n$ creation operator  takes
the matrix form: $\left[e^{in\theta} \right]_{k,k'}= \left[
\delta_{k,k'+n} \right]$, where $k=\ldots, -2,-1,0, 1,2, \ldots$ .
Truncating these infinite matrices with some large cutoff, one may
exponentiate the Hamiltonian and obtain the matrix form of
$\left[\hat {\cal U}_q\right]_{k,k'}$. The latter  may be
numerically diagonalized to find the function
$\epsilon^{(0)}_q(\alpha_m,\gamma)$. The free energy and the
transport barrier are then given by Eqs.~(\ref{free-energy}) and
(\ref{barrier}).

We start from the simplest case of section \ref{sec2}. The
monovalent salt  with the concentration $\alpha_{-1}=\alpha_{1}$
leads to the term $-\alpha_1 \delta_{k,k'+1}-\alpha_1
\delta_{k,k'-1}$ in the matrix $\left[\hat H_q\right]$. For
illustration we show the $4 \times 4$ truncation of $\left[\hat
H_q\right]$ and $\left[ e^{-i \theta} \right]$ matrices
$\left[\hat H_q\right] \rightarrow \left(\begin{array}{clcr}
(1+q)^2 & -\alpha_1 & 0 & 0  \\
-\alpha_1 & (0+q)^2 & -\alpha_1 & 0  \\
0 & -\alpha_1 & (-1+q)^2 & -\alpha_1  \\
0 & 0 & -\alpha_1 & (-2+q)^2 \\
\end{array}
\right)$ and $\left[ e^{-i \theta} \right] \rightarrow
\left(\begin{array}{clcr}
0 & 1 & 0 & 0  \\
0 & 0 & 1 & 0  \\
0 & 0 & 0 & 1  \\
0 & 0 & 0 & 0 \\
\end{array}
\right)$. For reasonable precision the truncation size of the
numerical calculation has to be much larger. Typically we used $40
\times 40$ and checked that a further increase does not affect the
results.

To calculate the ``energy'' band $F_q$ of a long channel for
certain $\alpha_1$ and $\gamma$ one can use the matrix form of
$\hat H_q$ and $e^{-i \theta} $, equate the largest eigenvalue of
$\hat {\cal U}_q$ in Eq. (\ref{one-period})
to $e^{-\epsilon^{(0)}_q/\gamma}$, and calculate
$F_q=k_BT\epsilon^{(0)}_{q} L/x_T$. Such a calculation gives the
transport barrier $U_L(\alpha_1,\gamma) \equiv F_{max} -F_{min}$
shown in Fig. \ref{figfgamma}.

The models in sections \ref{sec4} and \ref{sec5} are treated
similarly, except that they have salt ion terms $-{\alpha_2}
\delta_{k,k'+2}-2\alpha_2 \delta_{k,k'-1}$ and$-\alpha_1
\delta_{k,k'+1}-\alpha_2 \delta_{k,k'+2}-(\alpha_1+2\alpha_2)
\delta_{k,k'-1}$ respectively in $\left[\hat H_q\right]$. For the
model of section \ref{sec3}, $\left[\hat H_q\right]$  contains
salt ion term $-\alpha_1 \delta_{k,k'+1}-\alpha_1
\delta_{k,k'-1}$. However, $\hat {\cal U}_q$ is of the form of
Eq.~(\ref{uq2}) and its largest eigenvalue  is denoted as
$e^{-2\epsilon^{(0)}_q/\gamma}$.

\section{Effects of the finite length and the electric field escape}
\label{secxi}

The consideration of the previous sections was certainly an
idealization that neglected several important phenomena. The most
essential of them are: (i) the finite length $L$ of the channel;
(ii) the escape of the electric field lines from the water into
the media with smaller dielectric constant. Each of these
phenomena leads to a smearing of the ion-exchange phase
transitions transforming them into crossovers. The goal of this
section is to estimate the relative sharpness of these crossovers.

Consider first the effect of the finite length (still neglecting
the field escape). Close to the first order phase transition the
free energy admits two competing minima (typically at $q=0$ and
$q=1/2$) with the free energies $F_0(\alpha_1)$ and
$F_{1/2}(\alpha_1)$, see Fig.~\ref{qalternative} (we focus on the
alternating dopants example of section \ref{sec3}). Being an
extensive quantity, the free energy is proportional to the channel
length: $F_b\propto L$, where $b=0,1/2$. Each of these two minima
is characterized by a certain ion concentration
$n_b(\alpha_1)=-\alpha_1/(k_BTL)\partial F_b/\partial \alpha_1$.
In the vicinity of the phase transition at $\alpha=\alpha_c$ the
difference of the two free energies may be written as:
\begin{equation}\label{finiteL}
  \frac{F_0(\alpha_1)-F_{1/2}(\alpha_1)}{k_BT} =
   \Delta n_{\mbox{ion}} L \, \frac{\alpha_1-\alpha_c}{\alpha_c}\, ,
\end{equation}
where $\Delta n_{\mbox{ion}}=n_0(\alpha_c)-n_{1/2}(\alpha_c)$ is
the latent concentration of ions across the transition. Taking a
weighted sum of the two states, one finds that the  concentration
change across the transition is given by the ``Fermi function'':
\begin{equation}\label{crossover}
  \Delta n(\alpha_1)=\frac{\Delta n_{\mbox{ion}} }
  {e^{\Delta N(\alpha_c-\alpha_1)/\alpha_c}+1}\, ,
\end{equation}
where $\Delta N\equiv \Delta n_{\mbox{ion}}L$ is the total latent
amount of ions in the finite length channel. This gives for the
transition width $(\alpha_1-\alpha_c)/\alpha_c \propto 1/\Delta
N$. Therefore the transition is relatively sharp as long as
$\Delta N\gg 1$. For small enough $\gamma$ the number of ions
entering or leaving the channel at the phase transition is almost
equal to the number of dopants: $\Delta N\lesssim
N_{\mbox{dopants}}=\gamma L/x_T$. The necessary condition of
having a sharp transition, therefore, is to have many dopants
inside the channel. For example, for the transition of
Fig.~\ref{figlatentjump} $\Delta n_{\mbox{ion}}\approx
0.8\gamma/x_T$ and $\gamma=0.1$, one finds $\Delta N\approx 0.08
L/x_T $. At larger $\gamma$ the number of dopants increases (for
fixed length $L$), but the relative latent concentration is
rapidly decreasing, see Fig.~\ref{figlatentnumb}. Employing
Eq.~(\ref{gammac}), one estimates $\Delta N\approx \gamma
(1-2e^{-1/(4\gamma)})L/x_T$ this expression is maximized at
$\gamma \approx 0.15$, where $\Delta N \approx 0.1L/x_T$. We
notice, in passing, that even for $\Delta N\lesssim 1$, being
plotted as a function of $\log \alpha_2$, the function
(\ref{crossover}) still looks as a rather sharp crossover.

\begin{figure}[ht]
\begin{center}
\includegraphics[height=0.22\textheight]{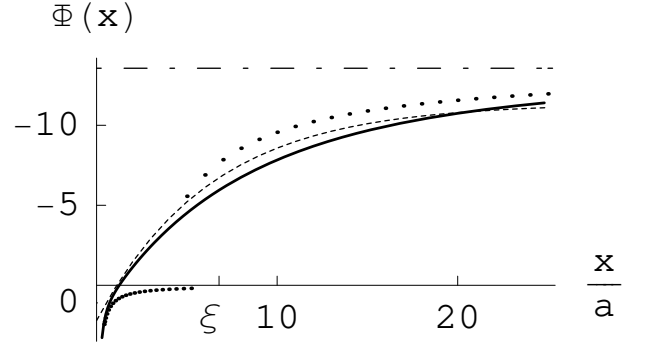}
\end{center}
\caption{Electrostatic potential $\Phi(x)$ in units of $e/\kappa_1
a$ of a point charge in an infinitely long channel as a function
of the dimensionless distance $x/a$ along the channel axis. Here
$a$ is the radius of the cylinder, and $\kappa_1/\kappa_2=40$. The
full line is an exact solution of the Laplace equation
\cite{Smythe,Parsegian}. The dotted lines are 3d Coulomb
potentials: the lower one is $-e/\kappa_1 x$; the upper one is
$\Phi_{\infty}-e/\kappa_2 x$, where $e\Phi_{\infty}=2 U_{\infty}$
and $U_{\infty}$ is the self-energy of a charge in the infinite
channel. The dashed line corresponds to Eq.\ref{xipotential}.
} \label{figsingle}
\end{figure}

We turn now to the discussion of the electric field  escape from
the channel, which happens due to the finite ratio of the
dielectric constants of the channel's interior, $\kappa_1$, and
exterior, $\kappa_2$. Fig.~\ref{figsingle} shows the electrostatic
potential of a unit point charge placed in the middle of the
channel with $\kappa_1/\kappa_2=40$, [\onlinecite{Smythe}]. The
potential interpolates between $-e/(\kappa_1 x)$ at small
distances, $x\lesssim a$, and $2U_\infty/e - e/\kappa_2 x$, where
$U_\infty = e^2\xi/(\kappa_1 a^2)$, at large distances $x>\xi$.
Here the length scale $\xi\simeq a\sqrt{\kappa_1/\kappa_2}\approx
6.8a$ is the characteristic escape length of the electric field
displacement from the interior of the channel into the surrounding
media with the smaller dielectric constant. The quantity $U_\infty
= U_L(0)2\xi/L$ is the excess self-energy of bringing a unit
charge inside the infinite channel.

In between the two limits the potential may be well approximated
by the following phenomenological expression (dashed line in
Fig.~\ref{figsingle}):
\begin{equation}\label{xipotential}
\Phi(x)=E_0\xi\left[1-e^{-|x|/\xi}-1.1\,{a\over\xi}\right]\, .
\end{equation}
Our previous considerations correspond to the limit $\xi\to
\infty$ (save for the last term). The last term in
Eq.~(\ref{xipotential}) originates from the fact that in the
immediate vicinity of the charge $|x|\lesssim a$ the electric
field is not disturbed by the presence of the channel walls. As a
result the length of $\approx 1.1 a$ is excluded from paying the
excess self-energy price. This leads to (typically slight)
renormalization of the effective concentrations $\alpha\to
\alpha_{eff}$. The more detailed discussion may be found in
Ref.~[\onlinecite{Kamenev}]; below we neglect the last term in
Eq.~(\ref{xipotential}).

One can repeat the derivation of section \ref{secanalytical} with
the potential Eq.~(\ref{xipotential}), employing the fact that
$\Phi^{-1}=(2E_0)^{-1}\delta(x-x')[\xi^2-\partial^2_x]$. As a
result, one arrives at Eq.~(\ref{Texponent}) with the modified
Hamiltonian $\hat H=(i\hat\partial_\theta)^2 -2\alpha_1
\cos\theta+ (x_T/2\xi)^2\theta^2$. Since the last term violates
the periodicity, the quasi-momentum $q$ is not conserved. However,
in the limit $x_T/\xi\ll 1$ one can develop the quasi-classical
approximation over this small parameter. Transforming the
Hamiltonian into the momentum representation, one notices that
$q(x)$ is a slow quasi-classical variable. As a result, the
partition function of the channel with the finite $\xi$ may be
written as:
\begin{equation}\label{slowq}
  Z=\!\int\!\! {\cal D}q(x)\, \exp \left\{ -\int\limits_0^L\!\! dx\left[ {\xi^2\over x_T}
  (\partial_x q(x))^2 + {1\over x_T}\, \epsilon_{q(x)}^{(0)} \right] \right\}\, ,
\end{equation}
where $F_q=k_BT \epsilon_{q}^{(0)}L/x_T $ is the free energy as
function of $q$  in $\xi \to \infty$ limit (no electric field
escape). This expression shows that there are no true phase
transitions even if $\epsilon_q^{(0)}$ possesses two separate
minima. Indeed, due to its finite rigidity  the $q(x)$ field may
form domain walls and wander between the two. As a result, the
first order transition is transformed into a crossover. Formally
Eq.~(\ref{slowq}) defines the ``quantum mechanics'' with  the
potential $\sim \epsilon_q^{(0)}$. The smearing of the transition
is equivalent to the avoiding crossing intersection due to
tunnelling between the two minima of the $\epsilon_q^{(0)}$
potential. Using this analogy, one finds for the concentration
change across the smeared transition:
\begin{equation}\label{crossover}
  \Delta n(\alpha_1)= \frac{\Delta n_{\mbox{ion}} }{2}\left[1+
  \frac{\alpha_1-\alpha_c}{\sqrt{(\alpha_1-\alpha_c )^2
+\alpha_c^2\delta^2}}\right]\, ,
\end{equation}
where $\delta$ is the WKB tunnelling exponent:
\begin{equation}\label{tunneling}
  \delta=\exp\left\{-{\xi\over x_T}\int\limits_0^{1/2}\!\!
  dq\, \sqrt{\epsilon_q^{(0)}-\epsilon_0^{(0)}}\right\}\, .
\end{equation}
As a reasonable approximation for $\epsilon_q^{(0)}(\alpha_c)$ one
may use (c.f. Eq.~(\ref{doublecos})) $\epsilon_q^{(0)}(\alpha_c)=
U_c/(8U_L(0)) \cos (4\pi q)$, where $U_c$ is the transport barrier
at the critical point. Substituting this expression in
Eq.~(\ref{tunneling}) one estimates $\delta\approx
\exp\{-\xi/(2\pi x_T)\sqrt{U_c/U_L(0)}\}$. Using $\xi\approx 6.8\,
a$ and $U_c\approx 0.2\,U_L(0)$ (cf. Fig.~\ref{figalternative}),
one obtains $\delta\approx \exp\{-l_B/a\}$. Thus for channels with
$a< l_B$ one obtains $\delta \lesssim 0.4$ and the crossover
Eq.~(\ref{crossover}) is relatively sharp.

\section{Contact (Donnan) potential}
\label{Donnan}

Until now we concentrated on the barrier proportional to the
channel length $L$ (or escape length $\xi$). If $\alpha \ll
\gamma$ there is an additional, independent on $L$, contribution
to the transport barrier. It is related to the large difference in
cation concentrations inside and outside the channel.
Corresponding contact (Donnan) potential $U_D$ is created by
double layers at each end of the channel consisting of one or more
uncompensated negative dopants and positive screening  charge near
the channel's mouth.

For $ \gamma \ll 1$ one finds $|U_{D}| \ll U_{L}(\gamma)$ and the
channel resistance remains exponentially large. When $\gamma$
grows the barrier $U_{L}(\gamma)$ decreases and becomes smaller
than $U_{D} =- k_B T \ln(\gamma/\alpha)$, which increases with
$\gamma$. In this case the measured resistance may be even smaller
than the naive geometrical diffusion resistance of the channel.

\begin{figure}[ht]
\begin{center}
\includegraphics[height=0.12\textheight]{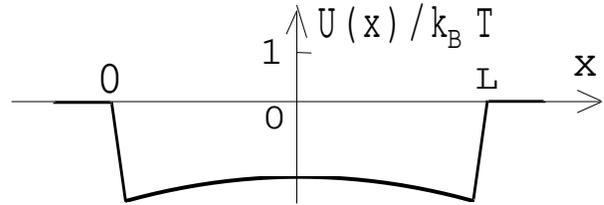}
\end{center}
\caption{The electrostatic potential for cations across the
channel with 5 dopants, considered in the text.}
\label{figschottky}
\end{figure}

Let us, for example, consider a channel with $L=5$ nm, $a = 0.7$
nm, $x_T=0.35$ nm at $c= 0.1$ M (which corresponds to
$\alpha=0.035$) and $\gamma=0.3$  (5 dopant charges in the
channel). The bare barrier $U_L(0) = 3.5 k_{B}T$ is reduced down
to $U_L(\gamma) = 0.2 k_{B}T$. At the same time $U_{D} = -
2.5k_{B}T$. Thus due to 5 wall charges, instead of the bare
parabolic barrier of Fig.~\ref{figshortchannel} we arrived  at the
wide well with the almost flat bottom (Fig.~\ref{figschottky}).

Unlike the $L$-dependent self-energy barrier considered above, the
role of the Donnan potential is charge sensitive. In the
negatively doped channel the Donnan potential well exists only for
the cations while anions see the additional potential barrier.
Thus, the Donnan potential naturally explains why in negatively
doped biological channels the cation permeability is substantially
larger than the anion one.

The contact potential $U_D$ may be augmented by the negative
surface charge of the lipid membrane~\cite{Appel} or by affinity
of internal walls to a selected ion,  due to ion--specific short
range interactions~\cite{Doyle,Hille}. It seems that biological
channels have evolved to compensate large electrostatic barrier by
combined effect of $U_{D}$ and short range potentials. Our theory
is helpful if one wants to study different components of the
barrier or modify a channel. In narrow artificial nanopores there
is no reason for compensation of electrostatic barrier. In this
case, our theory may be verified by titration of wall charges.

\section{Conclusion}
\label{secconclusion}

In this paper we studied the role of wall charges, which we call
dopants, played in the charge transport through ion channels and
nanopores. We considered various distributions of dopant charges
and  salt contents of the solution and showed that for all of them
doping reduces the electrostatic self-energy barrier for ion
transport. This conclusion is in qualitative agreement with
general statement on the role of transitional binding inside the
channel~\cite{Bezrukov}. In the simplest case of identical
monovalent dopants and a monovalent salt solution such a reduction
is monotonous and smooth function of the salt and dopant
concentrations. The phenomenon is similar to the low-temperature
Mott insulator-metal transition in doped semiconductors. However,
due to the inefficiency of screening in one-dimensional geometry
we arrived at a crossover rather than a transition even in
infinite channel.

A remarkable observation of this paper is that the interplay of
the ion entropy and the electrostatic energy  may lead to true
thermodynamic  ion--exchange  phase transitions. A necessary
condition for such a transition to take place is the competition
between more than one possible ground states. This in turn is
possible for compensated (e.g. alternating) doping or for mixture
of cations of various valency. The ion--exchange transitions are
characterized by latent concentrations of ions. In other words,
upon crossing a critical bulk concentration  a certain amount of
ions is suddenly absorbed or released by the channel. The phase
transitions also lead to non-monotonic dependencies of the
activation barrier as a function of the ion and dopant
concentrations. For simplicity we restricted ourselves with the
periodic arrangements of dopants. The existence of the phase
transitions is a generic feature based only on the possibility of
having more than one ground state with global charge neutrality.
Thus they exist for arbitrary positioned dopants. In reality the
phase transitions are smeared into relatively sharp crossovers due
to finite size effects along with the finite  electric field
escape length, $\xi$.

We have also demonstrated that the  doping can  make the channels
selective to one sign of monovalent salt ions or to divalent
cations. This helps to understand how biological K, Na, Ca
channels select cations and how Ca/Na channel selects Ca versus
Na. The surprising fact is that Ca$^{2+}$ ions, which could be
expected to have four times larger self-energy barrier, actually
exhibit  the same barrier as Na$^{+}$. This phenomenon is
explained by fractionalization of Ca$^{2+}$ on  two unit-charge
mobile solitons.

We study here only very simple models of a channel with charged
walls. This is the price for many asymptotically exact results.
Our results, of course, can not replace powerful numerical methods
used for description of specific biological channels~\cite{Roux}.

In the future this theory may be used in nano-engineering projects
such as modification of biological channels and design of long
artificial nanopores. Another possible nano-engineering
application deals with the transport of charged polymers through
biological or artificial channels. A polymer moves slowly and for
ions its charges may be considered as static. Therefore, for thin
and stiff polymers in the channel, the charges on polymers can
play the same role as doping. As a result,  all the above
discussions are directly applicable to the case of long charged
polymer slowly moving through the channel. Changing the polymer
one can change the dopants density.

In a more complicated scenario, the polymer  can be bulky and
occupy substantial part of the channel's cross-section. Important
example of such situation is translocation of a single stranded
DNA molecule through the $\alpha$-Hemolysin channel~\cite{Meller}.
In this case, the narrow part of the channel, immersed in the
lipid membrane ($\beta$-barrel) can be approximated as an empty
cylinder, while DNA may be considered as a coaxial cylinder
blocking approximately a half of the channel cross-section. The
dielectric constant of DNA is of the same order as one of lipids.
Thus, the electric field lines of a charge located in the water
gap between the two lipid cylinders are squeezed much stronger
than in the empty channel. This may explain strong  reduction of
the ion current in presence of the DNA, which is also different
for poly-A and poly-C DNA~\cite{Meller}.  The latter remarkable
observation inspires the hope that translocation of DNA may be
used as a fast method of DNA sequencing. We shall discuss the
bulky polymer situation in a future publication.

We are grateful to S. Bezrukov, A. I. Larkin and A. Parsegian for
interesting discussions. A.~K. is supported by the A.P. Sloan
foundation and the NSF grant DMR--0405212. B.~I.~S was supported
by NSF grant DMI-0210844.

\begin{appendix}

\section{Phase transitions at large dopant concentration}
\label{app1}

\begin{figure}[ht]
\begin{center}
\includegraphics[height=0.2\textheight]{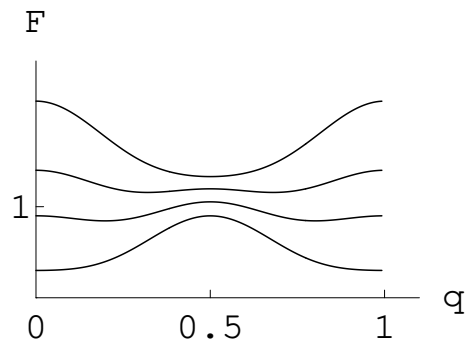}
\end{center}
\caption{Free energy in units of $10^{-10}U_L(0)$  as a function
of $q$ for $\gamma=1.7179330,~1.7179337,~1.7179341$ and
$1.7179346$ (from top to bottom) with the same $\alpha_2=5\cdot
10^{-5}$. The graphs are vertically offset for clarity. The ground
state continuously changes from $q=1/2$ to
$q=0$.}\label{figprocess}
\end{figure}

In this appendix we discuss some details of the phase transitions
at $\gamma_{c1}\approx 1.7$, $\gamma_{c2}\approx 5.3$,
$\gamma_{c3}\approx 10.8$, etc, visible in
Fig.~\ref{figdoubleabs}. The free energy $F_q$ as function of the
order parameter $q$ for a few values of $\gamma$ in the vicinity
of $\gamma_{c1}$ is plotted in Fig.~\ref{figprocess}. The minimum,
initially at $q=1/2$ for $\gamma_c<\gamma<\gamma_{c1}$, splits
into two minima symmetrical around $1/2$. These two gradually move
away from each other until they rich $q=0$ and $q=1$,
correspondingly, for $\gamma > \gamma_{c1}$.
The continuous variation of the absolute minimum (as opposed to
the discrete  a switch between the two fixed minima) suggests the
second order phase transition scenario, taking place at
$\gamma_{c1}$. The situation is more intricate, however. Namely,
there are two (and not one) very closely spaced second order
transitions, situated symmetrically around $\gamma_{c1}$ point. At
the first transition  the minima depart from $q=1/2$ and start
moving towards $q=0$ and $q=1$, while at the second one they
``stick'' to $q=0,1$. Therefore, unless one has a very fine
resolution (in $\gamma$ and/or $\alpha_2$), the entire behavior
looks as a single first order transition.

The $F_q$ functions of Fig.~\ref{figprocess} are very well  fitted
with the following phenomenological expression:
\begin{equation}\label{doublecos}
F_q(\gamma) = a (\gamma_{c1}-\gamma) \cos(2 \pi q) + b \cos(4 \pi
q)\, ,
\end{equation}
where $a \gg b > 0$. For any $\gamma$ the ground state corresponds
to the value $q_0$ which minimizes  $F_q$. Therefore $q_0$ is
found from: either $\cos (2 \pi
q_0)=(\gamma-\gamma_{c1}){a/(4b)}$, or $\sin(2\pi q_0)=0$. The
former equation has  solutions  only in the narrow interval
$\gamma_{c1}-4b/a<\gamma< \gamma_{c1}+4b/a $. In this interval of
dopant concentrations  the minima move from $q=1/2$ to $q=0,1$.
The edges of this interval constitute  two second-order phase
transitions located in the close proximity to each other. Near the
first transition $|q_0-1/2|\simeq \sqrt{\gamma
+4b/a-\gamma_{c1}}$, while near the second one: $|q_0|\simeq
\sqrt{\gamma -4b/a-\gamma_{c1}}$. Therefore the critical exponent
is the mean--field one: $\beta=1/2$. This could be anticipated for
the system with the long-range interactions.

It is interesting to notice that the first order transition,
discussed in the main text, may be also well fitted with
Eq.~(\ref{doublecos}) but with {\em negative} coefficients $a$ and
$b$. The accuracy of our calculations is not sufficient to
establish if the subsequent transitions are  the  first order ones
or pairs of the very closely spaced second order transitions. As
far as we can see the sequence of the reentrant transitions
continues at larger dopant concentrations. Notice, however, that
the difference of the corresponding free energies (and thus
associated latent concentrations) are exponentially small at large
$\gamma$.

\end{appendix}



\begin{thebibliography}{99}
\bibitem{ft6}The factor {$\alpha_2^2$ under the logarithm is
the result of having two ions at every dopant, and  factor of $6$
is obtained by integrating up the allowed volumes of the two ions
for all possible ion configurations around the dopant.}
\end{thebibliography}
\end{document}